\documentclass[journal]{IEEEtran}

\usepackage{amsmath, amsfonts, amssymb, cite, epsfig}
\usepackage{times, algorithm, algorithmic, amsbsy,bm,amsthm,graphicx}
\usepackage[justification=centering]{caption}
\usepackage{subfigure}
\usepackage{xcolor}

\newtheorem{theorem}{Theorem}

\newtheorem{definition}{Definition}

\usepackage{epstopdf}
\graphicspath{{./figures/}}

\def \FigWidthSSmaller{0.34\textwidth}
\def \FigWidthSmaller{0.37\textwidth}
\def \FigWidthSmall{0.42\textwidth}

\def\P{\mathbb P}
\def\E{\mathbb E}

\def\1{\mathbf 1}

\def\hlk{\color{black}}

\begin{document}
\title{How to Test the Randomness from the Wireless Channel for Security?}
%
%
%

\author{Zhe~Qu,~\IEEEmembership{Student~Member,~IEEE,}
        Shangqing~Zhao,~\IEEEmembership{Student~Member,~IEEE,}
        Jie~Xu,~\IEEEmembership{Member,~IEEE,}
        Zhuo~Lu,~\IEEEmembership{Senior~Member,~IEEE,}
        and~Yao~Liu,~\IEEEmembership{Senior~Member,~IEEE}
\thanks{Zhe Qu, Shangqing Zhao and Zhuo Lu are with the Department
of Electrical Engineering, University of South Florida, Tampa,
FL, 33620, USA.\protect\\
Emails: \{zhequ, shangqing, zhuolu\}@usf.edu}
\thanks{Jie Xu is with the Department of Electrical and Computer Engineering, University of Miami, Coral Gables, FL, 33146, USA.\protect\\ Email: jiexu@miami.edu}
\thanks{Yao Liu is with the Department of Computer Science and Engineering, University of South Florida, Tampa, FL, 33620, USA.\protect\\ Email: yliu@cse.usf.edu}}

\maketitle

\begin{abstract}
We revisit the traditional framework of wireless secret key generation, where two parties leverage the wireless channel randomness to establish a secret key. The essence in the framework is to quantify channel randomness into bit sequences for key generation. Conducting randomness tests on such bit sequences has been a common practice to provide the confidence to validate whether they are random. Interestingly, despite different settings in the tests, existing studies interpret the results the same: passing tests means that the bit sequences are indeed random.

In this paper, we investigate how to properly test the wireless channel randomness to ensure enough security strength and key generation efficiency. In particular, we define an adversary model that leverages the imperfect randomness of the wireless channel to search the generated key, and create a guideline to set up randomness testing and privacy amplification to eliminate security loss and achieve efficient key generation rate. We use theoretical analysis and comprehensive experiments to reveal that common practice misuses randomness testing and privacy amplification: (i) no security insurance of key strength, (ii) low efficiency of key generation rate. After revision by our guideline, security loss can be eliminated and key generation rate can be increased significantly.
\end{abstract}


\IEEEpeerreviewmaketitle

\section{Introduction}\label{sec:introduction}

Leveraging the wireless channel randomness has become one of the fundamental approaches to build low-cost security designs for emerging wireless applications, such as radio frequency identification (RFID) \cite{wang2018towards} and Internet of Things (IoT)  \cite{zhu2013extracting}. In particular, two communication parties, Alice and Bob, can use the random yet reciprocal wireless channel measurements, such as received signal strength information (RSSI) \cite{jana2009effectiveness, liu2012collaborative, 14pgk-seon}, channel state information (CSI) \cite{xi2016instant} and phase shifts \cite{wang2011fast, wang2012cooperative}, to generate a common secret sequence to build a security design, such as secret key generation \cite{mathur2008radio, jana2009effectiveness, wallace2010automatic, chen2011secret, liu2012collaborative, wang2011fast, wang2012cooperative, liu2013fast, tsouri2013threshold , zhu2013extracting, ali2014eliminating, 14pgk-seon, xi2016instant}, secure communication \cite{liang2008secure, liu2015secure} and user authentication \cite{argyraki2013creating}. Then, Alice and Bob can enter the cryptographic domain \cite{jana2009effectiveness, liu2012collaborative, 14pgk-seon} and use information reconciliation \cite{brassard1993secret} and privacy amplification \cite{bennett1995generalized, maurer2003secret} to compress their respective bit sequences. The framework is considered as secure enough because a third-party eavesdropper Eve is expected to gain little information about the shared secret if she is more than half the wavelength away from them because of wireless fading \cite{mathur2008radio}.


To evaluate the security of such a design, existing studies \cite{mathur2008radio, jana2009effectiveness,    wallace2010automatic, chen2011secret, liu2012collaborative, wang2011fast, wang2012cooperative, liu2013fast, tsouri2013threshold , zhu2013extracting, ali2014eliminating, 14pgk-seon, xi2016instant} adopt the NIST randomness tests \cite{nisttest} to evaluate whether the underlying secret bit sequences generated from the wireless channel exhibit randomness properties, which has become the common practice. {\hlk However, many of them \cite{mathur2008radio, jana2009effectiveness, liu2012collaborative, chen2011secret, xi2016instant} simply adopt the default NIST choices to set up a randomness test and consider successfully passing the test as a demonstration of security strength. Although passing a randomness test may hint that there is no major flaw in the design, it may not provide a guaranteed level of security stength. It becomes necessary to quantitatively understand the security impact of setting up randomness testing for the designs extracting random secrets from the wireless channel. }


At the same time, efficiency is always another important design aspect. In secret key generation \cite{jana2009effectiveness, liu2012collaborative, chen2011secret}, efficiency refers to the key generation rate, which depends on the strictness of the randomness test and the sequence compression rate for privacy amplification. However, there is no theoretical analysis in the literature on how to guarantee efficient secret key generation. Therefore, it is necessary to provide a design guideline for secrete key generation to ensure both security and efficiency.

In this paper, we ask a fundamental question: {\em \hlk How to properly set up statistical randomness tests for testing the wireless channel for both security and efficiency?}


In current study, the mismatched assumption between practical channel coherence and theoretical memoryless assumption leads to a gray area in realistic wireless key establishment applications. Each test in NIST assumes that the bit sequence is IID, but in practical wireless communication framework, the RSSI or CSI cannot be considered as IID \cite{mckay2005general, slepian1973noiseless}. Despite the correlation, it can also pass the NIST tests by choosing $\alpha = 0.01$ \cite{mathur2008radio, jana2009effectiveness, liu2012collaborative} and $0.05$ \cite{chen2011secret, xi2016instant}.

Before we verify the security level of any secret key generation design, the first step is to clearly define an adversary model. A formal adversary model will enable thorough security analysis and evaluation, but it has not been systematically studied in the literature. To this end, we study how an adversary can defeat a security design by taking advantage of a potential defect of the wireless channel, which is unpredictable with unknown ground truth. In particular, it is never known whether a channel can indeed yield independently random sequences for secret key generation. Therefore, under the assumption that the generated secret sequence is indeed statistically correlated, an adversary can search for it from the most-likely sequence candidate to the least-likely one, as opposed to random guessing. This strategy, called Maximum Likelihood Tree Search (MLTS), is considered for the security evaluation of a wireless channel randomness based design.


With the understanding of the adversary's capability, we propose a new design guideline for randomness testing, which involves solving an optimization problem that maximizes key generation efficiency under guaranteed security. In particular, we derive a mathematical formula for choosing the proper P-value threshold of nine different NIST tests. To our best knowledge, our design guideline is the first theoretical framework for wireless channel randomness based secret key generation, where the randomness test parameters are not empirically set. We note that rather than designing a new secret key generation method from the wireless channel, our focus is on how to properly setup the randomness tests. We conduct real-world experiments to validate the analysis, and incorporate our design guideline into seven popular key generation methods and compare the difference. The results show that (i) using our design guideline, these methods achieve zero security loss in various experiment scenarios; (ii) the key generation efficiency can be significantly improved; (iii) Our design guideline is more adaptive for generating different bit length of key sequences. In summary, the main contributions of this paper are as follows:
\begin{enumerate}
\item We {\hlk introduce the MLTS strategy} to formalize the security analysis and evaluation of wireless channel randomness based designs.

\item We propose a new design guideline on how to properly setup the randomness test parameter to eliminate security loss and achieve high efficiency.

\item We conduct the experiments in practical environments to show the improvement by our design guideline compared with existing secret key generation studies.
\end{enumerate}

\section{Background and Preliminaries}\label{Sec:Background}
In this section, we briefly introduce the background of extracting secret from the wireless channel, and then formalize the framework of secret key generation. To this end, we discuss the scenario and assumptions in this paper.
\subsection{Extracting Random Secrets from Wireless Channels}
Traditional cryptographic mechanisms (e.g., Diffie-Hellman and RSA \cite{kaufman2002network}) reply on establishing computational difficulties for an adversary to achieve the goal of security. In wireless, mobile or IoT domains \cite{zhu2013extracting, xi2016instant}, many wireless security designs have been proposed to leverage the reciprocal and random properties of wireless channel measurements (e.g., RSSI and phase shifts) to generate a common secret sequence between Alice and Bob. In many studies \cite{azimi2007robust, mathur2008radio, jana2009effectiveness, patwari2010high, wallace2010automatic, chen2011secret, liu2012collaborative, wang2011fast, wang2012cooperative, liu2013fast, tsouri2013threshold, wei2013adaptive, zhu2013extracting, ali2014eliminating, 14pgk-seon, ponnaluri2016practical, 5492690}, such a sequence is directly used as the secret key for the secure communication between Alice and Bob. In this paper, we use secret key generation as our main application scenario to study how to test the wireless channel randomness for security, since it is the most representative study of security designs leveraging wireless channel randomness.

\begin{figure}[t!]
\begin{center}
\includegraphics[width=\FigWidthSmaller]{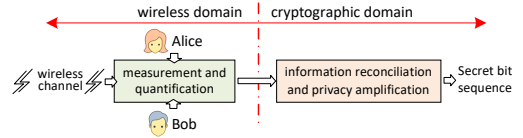}
\caption{Typical framework of random secret key generation from wireless channel between Alice and Bob.}
\label{Fig:KeyEstFramework}
\end{center}
\end{figure}

Figure~\ref{Fig:KeyEstFramework} shows that a typical framework for Alice and Bob extracting a common secret sequence from the wireless channel. The framework consists of design components in both wireless domain and cryptographic domain. In the wireless domain (shown on the left-hand side of Figure~\ref{Fig:KeyEstFramework}), Alice and Bob keep measuring the wireless channel response, such as measuring the RSSI, CSI or phase shifts between them, and then quantify the measurements into bits \cite{azimi2007robust, mathur2008radio, jana2009effectiveness, wang2012cooperative, liu2012collaborative, zhu2013extracting}. Because of the reciprocal property of the wireless channel, their measurements are likely to be the same from the channel between them, and accordingly their quantified bits sequences should also be likely the same.

Then, Alice and Bob can enter the cryptographic domain \cite{jana2009effectiveness,  liu2012collaborative, xi2016instant} as shown on the right-hand side of Figure~\ref{Fig:KeyEstFramework} and use information reconciliation \cite{brassard1993secret} and privacy amplification \cite{bennett1995generalized, maurer2003secret} to compress their respective bit sequences (with low per-bit entropy) into the same short sequence (with per-bit entropy expected to be near 1).

\subsection{Formalizing the Framework of Secret Key Generation}\label{SubSec:Model}
For the framework to extract random bit sequences from the wireless channel, there are two major components between Alice and Bob: the wireless domain design and the cryptographic domain design, which are formally modeled in the following.

\begin{definition}[Secret Bit Sequence Extraction Models]\label{Def:KeyGen}
~\\
\vspace{-0.5cm}
\begin{itemize}
\item The wireless domain design is a function
\begin{equation}\label{Eq:f_w}
f_w: \Omega_D \rightarrow \{0,1\}^L
\end{equation}
mapping a random channel property (e.g., RSSI or phase shifts) in the continuous domain $\Omega_D$ during a time duration $D$ to a binary bit sequence in $\{0,1\}^L$, which denotes the set of all bit sequences with length $L$.

\item The cryptographic domain design is a function
\begin{equation}
f_c: \{0,1\}^L \rightarrow \{0,1\}^M
\end{equation}
mapping a binary bit sequence with length $L$ to a new sequence with shorter length $M \leq L$, in which the correlation among bits is minimized close to 0 by privacy amplification. When $M=L$, there is no cryptographic domain in a secret key generation design \cite{mathur2008radio, ali2014eliminating}, we simply set function $f_c$ as $f_c(x) = x$ for any input $x$.

\item A statistical randomness test is a function
\begin{equation}\label{Eq:S}
T: \{0,1\}^* \rightarrow \{\{\text{Accept $H_0$}\},\{\text{Accept $H_1$}\}\},
\end{equation}
where $\{0,1\}^*$ denotes the set of bit sequences with any length {\hlk (e.g., length of $L$ or $M$)}, $H_0$ and $H_1$ are null and alternative hypotheses denoting the events that the randomness test succeeds and fails, respectively.
\renewcommand{\footnotesize}{\scriptsize}
The objective of Alice and Bob is to leverage the random channel property between them, denoted by $\omega_D \in \Omega_D$\footnote{Alice and Bob may not observe exactly the same $\omega_D$ in practice because of noise and interference. In this regard, denote by $\omega_D^A$ and $\omega_D^B$ Alice's and Bob's observations respectively. Robust wireless domain design aims to achieve $f_w(\omega_D^A) = f_w(\omega_D^B)$ and information reconciliation also ensures $f_c(f_w(\omega_D^A)) = f_c(f_w(\omega_D^B))$. So $\omega_D^A \neq \omega_T^B$ does not affect our security analysis. For the sake of simple notation, we let $\omega_T = \omega_T^A = \omega_T^B$.} during time period $D$, to compute a bit sequence
\begin{equation}\label{Eq:K_T}
K_D = f_c(f_w(\omega_D)) \in \{0, 1\}^M
\end{equation}
for their security design purpose.
\end{itemize}
\end{definition}

In the extraction models, there is no evaluation that the bit sequence $K_D$ is sufficient for the security purpose. Therefore, security evaluation is another critical component for any wireless channel randomness based security design. To this end, NIST statistical randomness test suite \cite{nisttest} is widely adopted as a common practice in the literature \cite{mathur2008radio, jana2009effectiveness, wang2012cooperative} to test whether the generated bit sequence is random for the security purpose.

\subsection{Testing Randomness from Wireless Channels}\label{SubSec:TestingRandomness}
The procedure of a randomness test in the NIST test suite \cite{nisttest} to test the bit sequences extracted from the wireless channel is straightforward: for a bit sequence $X$ quantified from the wireless channel, compute the statistics of $X$ based on a particular test, called P-value, and compare this P-value with a threshold $\alpha$. The test succeeds if the P-value is greater than $\alpha$, and fails otherwise.

For Alice and Bob, failing the NIST tests indicates that the wireless channel measurement does not have enough randomness \cite{jana2009effectiveness}. They have to wait for a better channel condition or adjust their design parameters and then test again. Thus, randomness tests serve a critical role in evaluating the security of a design leveraging wireless channel randomness.

At first glance, it seems perfectly fine to use randomness tests for security evaluation because they are recommended for cryptographic use. But the key question here is not why, but how to use them to test the wireless channel randomness for security? We notice that existing studies adopt statistical randomness tests in different ways for security evaluations. Particularly, two major discrepancies exist in the literature.

\begin{enumerate}
\item Where to set the randomness test? There indeed exists a discrepancy in the literature to place the randomness test: a large number of designs \cite{jana2009effectiveness, 14pgk-seon, liu2012collaborative, liu2013fast} choose to test the bit sequences at Position~2, and other designs test at Position~1 \cite{mathur2008radio, ali2014eliminating, chen2011secret}.

\item How to choose a critical parameter, the P-value threshold $\alpha$, in randomness tests? The value of $\alpha$ represents the confidence level of the test output. We notice that $\alpha=1\%$ is dominantly adopted in existing studies \cite{mathur2008radio, ali2014eliminating, jana2009effectiveness, 14pgk-seon, liu2012collaborative} according to the NIST test suite \cite{nisttest}.  However, some studies also choose $\alpha=5\%$ for the tests, for example, \cite{xi2016instant} tests 200-bit sequences generated from the wireless channel between mobile devices.
\end{enumerate}

\begin{figure}[t!]
\begin{center}
\includegraphics[width=\FigWidthSmaller]{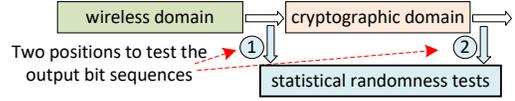}
\caption{The use of statistical testing at different positions in secret key generation from the wireless channel.}
\label{Fig:StTestPositions}
\end{center}
\end{figure}

{\hlk These observations in fact reveal that despite the advance on efficiently quantifying the wireless channel randomness (e.g., RSSI or CSI) into bit sequences, the common practice of using randomness testing exhibits discrepant setups for security evaluation. Are these setups equally secure, or secure enough for a particular application? As a result, the focus of this paper is to design a rigorous mechanism to understand how to properly use randomness tests for security evaluation of these security designs quantifying wireless channel randomness.}




\section{Problem Formulation and Research Statement}
We first formalize the models in extracting secret bit sequences from the wireless channel, then identify research challenges and propose the guideline for setting up statistical randomness tests.

\subsection{Formalizing the Role of Statistical Randomness Testing for Security}\label{SubSec:Models}

It is clear in Definition~\ref{Def:KeyGen} that extracting a secret bit sequence from \eqref{Eq:K_T} does not rely on the statistical randomness test $T$ in \eqref{Eq:S}. The role of $T$ is to ensure {\it security by denial}:  if the bit sequence from the channel fails the test $T$, then the channel randomness is not sufficient for the security purpose.

However, a randomness test can be set up in many different ways (e.g., varying setups observed in the literature: \cite{mathur2008radio, jana2009effectiveness, liu2012collaborative} vs \cite{chen2011secret, xi2016instant}). Due to the fact that a randomness test only considers some specific parts to evaluate the degree of randomness for a bit sequence, the bit sequence can always pass a randomness test as long as its construction is biased toward the test. Blindly setting up the randomness tests provides no guarantee of security. We have to rethink about how to quantify the extent to which the security by denial via a randomness test meets the security goal of Alice and Bob, i.e., obtaining a secret bit sequence $K_D$ of length $M$ in \eqref{Eq:K_T}.

To this end, we need to design $T$ in two steps: (1) Designing a technical adversary model Eve against Alice and Bob. However, there is no such model proposed in the literature, which makes formal analysis for wireless randomness based security difficult. (2) Defining Eve's attack success probability as a function of the randomness test. In this way, we can quantitatively measure the benefit of security by denial via a randomness test and properly set the test.

If the two steps are in place, we are able to evaluate whether a randomness test is properly set up for protecting the system security. Specifically, we aim to compare Eve's strategy under the randomness test with the benchmark random guess (RG) strategy in terms of the success probability, and set up the randomness test such that
\begin{equation}\label{Eq:EveBasicGuideline}
\P (\text{Eve succeeds}) \leq \P(\text{RG succeeds}).
\end{equation}

In other words, we must set up the randomness test such that Eve's attack is no better than RG. We also define the security loss due to the randomness testing as
\begin{equation}\label{Eq:BasicGuideline}
L_{\text{security}} = \log_2 (\P (\text{Eve succeeds}) / \P(\text{RG succeeds})).
\end{equation}
For example, if $\P(\text{Eve succeeds}) = 2^{-80}$ and $\P(\text{RG succeeds})$  $= 2^{-120}$, the security loss $L_{\text{security}}$ is computed as 40, which is the difference between the exponents in the two probabilities. In general, the security loss $L_{\text{security}}$ can have a negative value for some naive attack strategy (e.g., Eve always tries a fixed guess every time, which is even worse than RG). In this paper, we only consider non-trivial cases with $L_{\text{security}} \geq 0 $
(i.e., a security loss is non-negative).

\subsection{Formalizing the Role of Statistical Randomness Testing for Efficiency}\label{SubSec:Keyestablish}
In Section~\ref{SubSec:Models}, we formalize the role of statistical randomness testing from the security perspective. It is also worth noting that efficiency for secret key generation is another important factor to consider. Note that the efficiency comes from two aspects: randomness test $T$ (i.e., the probability of bit sequence can be accepted or rejected) and privacy amplification rate $R_{\text{privacy}}$ (i.e., the compressed rate $M/L$). If statistical randomness test $T$ is set too strict, it will be difficult to generate wireless secrets during a short time period because $T$ rejects the channel samples too many times. On the other hand, privacy amplification allows the input of a correlated bit sequence and compresses it into a short sequence with higher per-bit entropy. A higher $R_{\text{privacy}}$ can loose the design of the test $T$, but at the same time reduce the efficiency because more bits are compressed, which indicates more bits extracted from wireless channel are discarded. In this paper, we only aim to achieve high efficiency by controlling $T$ and $R_{\text{privacy}}$. Thus, we consider the efficiency $E$ as our evaluation and formally define it as
\begin{equation}\label{Eq:keyestablish}
E = \P(T~\text{accepts}~H_0 ) \cdot R_{\text{privacy}},
\end{equation}
and the efficiency loss is defined as $L_{\text{efficiency}} = 1 - E$. {\hlk Note that there are two situations of placing randomness test $T$ in the literature: if we place randomness test $T$ at position 1, $\P(T~\text{accepts}~H_0) = \P(T(f_w (\omega_D ))~\text{accepts}~H_0)$. If $T$ is at position 2, $\P(T~\text{accepts}~H_0 )=\P(T(K_D)~\text{accepts}~H_0)$. For simplicity, we generalize these two situations into one formula $\P(T~\text{accepts}~H_0)$.} With security loss $L_\text{security}$ and efficiency loss $L_\text{efficiency}$, we can quantitatively evaluate a wireless secret bit generation design in terms of its security and efficiency. To properly set up the randomness testing, we must ensure that its security loss $L_\text{security}$ is zero under an adversary model and at the same time we also need to maximize its efficiency $E$ or, equivalently, minimize the efficiency loss $L_\text{efficiency}$.



\section{Formal Adversary Model}

With the definition of security loss and efficiency, our next goal is to define a formal adversary model to measure the security loss. It is well known that the wireless channel response is statistically correlated over time, but two channel response samples with interval larger than the coherence time are approximately independent \cite{goldsmith2005wireless}. This approximate independence assumption has been widely used for wireless communication performance analysis and evaluation. However, whether this assumption is able to serve a basis for security design is not fully investigated. We aim at defining an adversary model that takes advantage of imperfect channel independence to launch attacks targeting the secret random bit generation framework.

We first present the scenarios and assumptions, then model the secret bit extraction from the wireless channel, and finally propose and analyze the adversary model.

\subsection{Scenarios and Assumptions}\label{SubSec:Assumptions}
We consider a wireless channel randomness based design scenario shown in Figure~\ref{Fig:StTestPositions}. We assume that all design specifications and parameters in the wireless domain {\hlk (e.g., bandwidth, carrier frequency and quantization methods)} and the cryptographic domain {\hlk (e.g., cryptographic methods)} in Figure~\ref{Fig:StTestPositions} are known to the public. An adversary Eve can hear all communications between Alice and Bob, but cannot access Alice's or Bob's antenna. Therefore, she cannot directly measure the accurate channel response between Alice and Bob. We assume that Eve can neither actively affect the wireless channel between Alice and Bob, nor modify the content of any communication.

We also assume that Eve has a powerful yet finite computational capability. This enables Eve to perform intensive computations, leveraging the imperfect randomness of the wireless channel measurements, to search for the secret between Alice and Bob. Such an assumption of Eve's practical computational capability helps offer intuitive measurements of security degradation to indicate the importance of correctly setting up statistical testing for wireless security. For example, the 2017 SHA-1 collision attack has an estimated computational effort equivalent to $2^{63.1}$ SHA-1 calls \cite{sha1attack2017}. We define Eve's capability and objective as follows.

\begin{definition}[Eve's Capability and Objective]\label{Def:Attacker} Given Definition~\ref{Def:KeyGen}, Eve aims to develop a key search strategy to maximize her success probability by performing $N$ searches for the secret $K_D$, and $N$ is called Eve's capability.
\end{definition}

\subsection{Analyzing Secrets Generated from Wireless Channel Randomness}

Given the fact that all models in Definition~\ref{Def:KeyGen} are publicly known, the objective of Eve is to develop a strategy to efficiently search for $K_D$ without exact knowledge of Alice and Bob's channel response $\omega_D$. Existing studies have well explored the building of functions $f_w$ and $f_c$ to obtain a key from \eqref{Eq:K_T}, but never fully investigated Eve's strategy. Suppose Eve has no smarter strategy but RG, if we assume that she has a maximum capability $2^{64}$, the probability is $2^{64} / 2^{128} = 2^{-{64}}$ to obtain a 128-bit key generated by Alice and Bob.

Is there a smarter strategy for Eve to do better? We first analyze how the wireless channel generates secret bit sequences. \cite{mathur2008radio} proposed the basic idea of the level-crossing algorithm: the channel information (e.g., RSSI or CSI) is estimated over a time interval (TI) larger than the coherence time, set two thresholds $q_+$ and $q_-$ by calculating magnitude or phase; the measured channel information will be quantified to 1, if it is greater than a threshold $q_+$ or 0 less than $q_-$. {\hlk It is widely suggested in existing works \cite{mathur2008radio, jana2009effectiveness, zhu2013extracting} that the measurement interval should at least equal the channel coherence time such that the measured samples are considered approximately independent.}




{\hlk Nonetheless, this approximate independence assumption creates a very vague boundary from the security perspective: the measurements from wireless channel are correlated \cite{mathur2008radio, jana2009effectiveness, zhu2013extracting, kim2018channel}; although the correlation becomes weaker and could be considered approximately independent for traditional performance analysis when the measurement interval increases, it still makes sense for security analysis to assume that the output bits from the wireless domain are statistically correlated rather than approximately independent. In other words, the input of function $f_w$ in \eqref{Eq:f_w} is regarded as a correlated signal, leading to a correlated output model of $f_w$.}

\begin{definition}[Wireless Bit Generation Model]\label{Def:BitGen}
~\\
Given a channel measurement and quantification period $D$, the output from the wireless domain, denoted as bit sequence $X = f_w(\omega_D) = [x_1, x_2, \cdots, x_L]$, is modeled as a binary correlated sequence with correlation coefficient $\rho \in [-1, 1] $ for consecutive bits $x_i$ and $x_{i+1}$ for $i\in [1, L-1]$, which is written as
\begin{equation}\label{Eq:rho}
\rho = \frac{\emph{\text{cov}}(x_i, x_{i+1})}{\sigma(x_i) \sigma(x_{i+1})},
\end{equation}
where $\text{cov}(x_i, x_{i+1})= \E \left( (x_i - \E(x_i)) (x_{i+1} - \E(x_{x+1})) \right)$ is the covariance between $x_i$ and $x_{i+1}$, and $\sigma(x_i)^2 = \E ((x_i - \E(x_i))^2) $ is the standard deviation of $x_i$.
\end{definition}

Definition~\ref{Def:BitGen} offers a more practical and generic model compared with the traditional one that assumes that channel samples are approximately independent with the sampling duration larger than the coherence time used in the literature. Apparently, we can set $\rho=0$ to obtain from the correlated model to the traditional one. Moreover, from a security perspective, we should always assume a defective (rather than perfect) randomness model for security design. A good wireless domain design should generate a bit sequence with a correlation coefficient $\rho$ close to 0. But we can never know a design indeed achieves 0 in practice. Thus, it is always good to assume $|\rho|>0$ even it is a very small value. Accordingly, Eve can leverage such a  model to construct her attack strategy, which in turn facilitates formal security analysis for statistically testing wireless channel randomness.

\begin{figure}
   \centering
   \includegraphics[width=\FigWidthSmall]{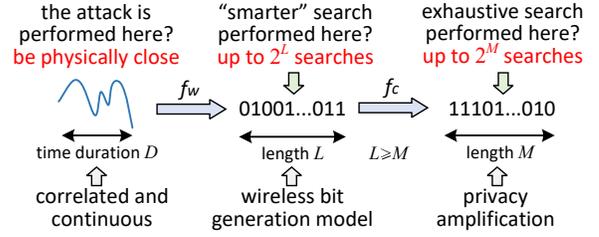}
   \caption{Eve's perspective on the secret key generation.}
   \label{Fig:AttackMotivation}
\end{figure}

\subsection{Eve's Strategy}
Given the bit generation model from the wireless domain, let us look at the secret bit generation from Eve's perspective, shown in Figure~\ref{Fig:AttackMotivation}. As Eve knows $K_D = f_c(f_w(\omega_D))$ from \eqref{Eq:K_T}, there are three straightforward strategies.
\begin{enumerate}

\item Search for the secret $K_D$ in the $\{0, 1\}^M$ space. There is no evident strategy better than RG because the last step of $f_c$ is privacy amplification. 

\item Search for the wireless domain output $X=f_w(\omega_D)$ in the $\{0, 1\}^L$ space. Then, compute $K_D=f_c(X)$ because $f_c$ is public. Note that $L \geq M$, so is it really worth searching in a potentially larger space? We find that leveraging the bit correlation to search for $X$ in $\{0, 1\}^L$ can result in a better success probability than RG in $\{0, 1\}^M$.

\item Search for $\omega_D$, then compute $K_D$ = $f_c(f_w(\omega_D))$. We note that this is possible only if Eve can physically access Alice's or Bob's antenna. We assume that Eve has no such access in this paper.
\end{enumerate}

\begin{figure}[ht]
   \centering
   \includegraphics[width=\FigWidthSSmaller]{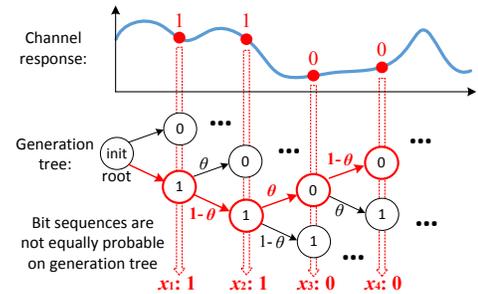}
   \caption{Bit generations from the wireless domain forms a generation tree.}
   \label{Fig:AttackGenerationTree}
\end{figure}

In the three strategies, we show that the second one for Eve (i.e., searching for $X=f_w(\omega_D)$ then computing $K_D=f_c(X)$) can generate higher success probability if Eve leverages the correlation in the wireless bit generation model in Definition~\ref{Def:BitGen}. Figure~\ref{Fig:AttackGenerationTree} illustrates an example of how the first 4 bits $x_1$, $x_2$, $x_3$ and $x_4$ from $X$ are generated: the wireless channel is slowly varying and the wireless domain design samples and quantifies the channel response into bits in a sequential way. The first two bits $x_1$ and $x_2$ are 1 and the channel changes so the last two bits $x_3$ and $x_4$ are 0.

According to Definition~\ref{Def:BitGen}, we map the wireless-domain generation into an abstract process in a generation tree as shown in Figure~\ref{Fig:AttackGenerationTree}, which enumerates all possible bit values generated sequentially. A path from the root to a leaf can represent a generated bit sequence. For example, Figure~\ref{Fig:AttackGenerationTree} shows the path that generates 1100. We denote by $\theta$ the value transition probability in the tree (i.e., the probability that the values of two consecutive bits are different). When the correlation coefficient is larger than 0, the correlation among bits in fact means that a generated bit is more likely to have the same value of the previous generated bit, i.e., $\theta$ should be smaller than 0.5. If the correlation coefficient is smaller than 0, $\theta$ should be larger than 0.5, respectively. As a consequence, all paths from the root to leaves in the tree exhibit different probabilities. This helps Eve because she can search for $X$ by starting from the most likely bit sequence towards the least likely bit sequence in the tree. We call such a strategy maximum-likelihood tree search (MLTS). 



MLTS maximizes Eve's success probability if bits in $X$ are statistically correlated (i.e. $|\rho| > 0$), and has equal performance to RG otherwise (e.g., $\rho=0$). In the following, we show the attack performance of MLTS.

\begin{theorem}[Maximum-Likelihood Tree Search]\label{TM:MLTS}
~\\
For the sake of simple notation, we let Eve's computational capability $N$ in Definition~\ref{Def:Attacker} satisfy $N=\sum_{i=0}^{n/2} { L \choose i} + \sum_{j=0}^{n/2} {L \choose j}$, where $0 \leq n \leq L$. Then, given the fact that a secret $K_D$ has been established, the attack success probability of \emph{MLTS} is
\begin{eqnarray}\label{Eq:beta}
&&\P(\text{\emph{MLTS succeeds}} \, |\, \text{\emph{$K_D$ established}})\\
&& = I_{\frac{1-\rho}{2}}(L-\frac{n}{2}, \frac{n}{2}+1) + I_{\frac{1+\rho}{2}}(L-\frac{n}{2}, \frac{n}{2}+1) = I_{\text{MLTS}}\nonumber,
\end{eqnarray}
where $I_x(a,b)$ is the regularized incomplete beta function
\begin{equation}\label{Eq:U_xab}
I_x(a,b) = \frac{B(x;a,b)}{B(a,b)}
\end{equation}
with incomplete beta function $B(x;a,b) = \int_0^x t^{a-1}(1-t)^{b-1} dt$ and complete beta function $B(a,b)=B(1;a,b)$.
\end{theorem}

{\noindent\it Proof:} To facilitate smooth presentation of our design and results, we defer the proof to Appendix~\ref{Sec:ProofTheorem} for details. \hfill~$\Box$

The advantage of MLTS is that it does not need to know the value of $\rho$ and the transition probability $\theta$ to work. Eve should always try to use MLTS in practice to search for $X$ then compute $K_D$. It is worth noting that we use the number of searches as an indicator of computational complexity. We consider Eve performs one search on a sequence if Eve spends some computations on the sequence. Due to the use of randomness testing and the use of hundreds of bits as a key in today's practice, Eve cannot easily exclude a wide range of sequences of hundreds of bits (or easily prone a large branch of the search tree) during searching for the correct key. Eve has to test sequences one by one. Even for a bit sequence that fails the randomness test, she still has to test it (thereby spending some computational time) before knowing that it cannot be used as the key. Or Eve can skip the test and directly spend computations on verifying if a key candidate is correct. These computations on the bit sequence constitute one search regardless of the fact that the bit sequence is test-compliant or not. Therefore, there is no straightforward way to skip all the sequences that are not test-compliant. Knowing the fact that $K_D$ is established does not reduce the number of searches to be performed by Eve. Note that we do not consider trivial cases here (e.g., Eve can simply exclude all 0 or 1 bits).

\hlk

\begin{figure}
   \centering
   \includegraphics[width=\FigWidthSmall]{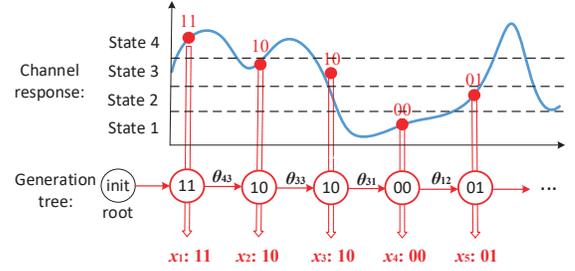}
   \caption{Eve's perspective on the secret key generation of 4-level quantization.}
   \label{Fig:attackmulti}
\end{figure}

In order to improve the efficiency of key generation, multi-level quantization methods have been developed in \cite{yasukawa2008secret, zeng2010exploiting}. 
The core of MLTS, which is to search from the most likely sequence to the least likely sequence, can also been applied to multiple level quantization. Fig.~\ref{Fig:attackmulti} shows an example of 4-level quantization of the channel response. The quantization includes 4 states (i.e., 11, 10, 01, 00). As the figure shows, the system first quantifies the channel response to state 4 (11), then state 3 (10), state 3 (10), state 1 (00), and state 2 (01), which leads to the sequence of 1110100001. The correlation of two states is stronger if they are closer. As a result, Eve's MLTS can be executed from the most likely sequence to the least likely sequence on a tree with multiple-bit states (instead of single-bit ones) as individual nodes. In this paper, we will focus on the single-bit quantization as it has been widely used in existing studies \cite{mathur2008radio, jana2009effectiveness, wallace2010automatic, chen2011secret, liu2012collaborative, 14pgk-seon, xi2016instant}. We provide basic performance analysis of MLTS for multiple-level quantization in Appendix~\ref{Sec:multi}.

\hlk


\section{Guidelines for Statistical Randomness Test Settings}
With the clearly defined MLTS-based attack model for Eve, we are ready to address how Alice and Bob should test the wireless channel randomness for security. Alice and Bob must make sure that they will not create a common secret from the wireless channel with a high correlation over time. On the other hand, they never know the exact value of channel correlation in practice. Then, it seems natural for them to test the channel first, and then make a binary decision, which formalizes the role of the randomness test $T$ in Definition~\ref{Def:KeyGen}.

In this section, we analyze Eve's success probability as a function of randomness testing, then propose the guideline for Alice and Bob to properly set up the randomness testing for security and efficiency.

\subsection{Eve's Success Probability as A Function of Randomness Testing}
{\hlk Randomness testing aims to eliminate the security loss defined in \eqref{Eq:BasicGuideline}} by denial and we should make sure that Eve's success probability is no better than the RG's success probability (i.e., no security loss). Based on Definition~\ref{Def:KeyGen}, we define $\P(\text{Eve succeeds})$ as
\begin{eqnarray}\label{Eq:Keygoal}
\P(\text{Eve succeeds}) && = \P(T~\text{Accept $H_0$}) \P (\text{Eve succeeds}|H_0)\nonumber\\
&& \leq \P(\text{RG succeeds}),
\end{eqnarray}
where
\begin{itemize}
\item $\P(\text{Accept $H_0$})$ is the probability that randomness test $T$ passes, which depends on the settings of $T$.
\item $\P(\text{RG succeeds})=N2^{-M}$ with Eve's capability $N$ and key length $M$.
\item $\P (\text{Eve succeeds}|H_0) \approx I_{\text{MLTS}}$ obtained in \eqref{Eq:beta}, denoting the probability that Eve obtains the key $\Omega_D$ conditioned on randomness test $T$ passes\footnote{{\hlk Eve aims to search the wireless domain output $X$ yielding the key $K_D$. However, due to hash collision in privacy amplification, there exists a probability $\P(\text{collision})$ that Eve finds another bit sequence $X'\neq X$ satisfying $K_D \!=\! f_c(X) \!=\! f_c(X')$. As a result, $\P(\text{Eve succeeds}) \!=\! I_{\text{MLTS}} \!+\! \P(\text{collision})$,} {\hlk where $\P(\text{collision})$ can be approximated as $1-(1-2^{-M})^L$ in \cite{henson2003analysis}.} {\hlk Since privacy amplification $f_c$ is always designed to make the collision probability $\P(\text{collision})$ negligible,} {\hlk for example, $L=M=32$ and $N=16$, $I_{\text{MLTS}} \geq \P(\text{RG}) = 1.53\times 10^{-5}$ and $\P(\text{collision})=7.45\times10^{-9}$ such that $I_{\text{MLTS}} \gg \P(\text{collision})$.} {\hlk Therefore, we approximate that $\P(\text{Eve succeeds}) \approx I_{\text{MLTS}}$ for a sufficiently large capability $N$ for Eve in this paper.}}.
\end{itemize}


According to our analysis, Eve should always use MLTS in practice and hope for a large $|\rho|$. As a result, Eve's success probability can be written as
\begin{eqnarray}\label{Eq:evesucceeds}
&& \P(\text{Eve succeeds}) = \nonumber\\
&& \P(\text{MLTS succeeds} \,|\, \text{$K_D$ established}) \P(\text{$K_D$ established}) \nonumber\\
&& = I_{\text{MLTS}} \cdot \P(\text{$T$ accepts $H_0$}) \leq \P(\text{RG succeeds}), \label{Eq:IP(H_0)}
\end{eqnarray}
where the last equality follows from Theorem~\ref{TM:MLTS}. Due to the fact that the binary sequence can be considered random enough after privacy amplification, MLTS only can focus on the sequence in the wireless domain. The more random (randomness test can reject large $|\rho|$ scenarios) and longer binary sequence ($L \geq M$) could eliminate the security loss.

\subsection{Observations and Design Guideline}\label{SubSec:Guideline}


From \eqref{Eq:IP(H_0)}, we can only guarantee that there is no security loss but the test $T$ might be set too strict to cause a lower efficiency $E$. Hence, efficiency $E$ is also important and should be considered at the same time of the design guideline. As a result, the design guideline is proposed to find the settings for test $T$ and the privacy amplification rate $R_{\text{privacy}}$ to maximize efficiency under the constraint of no security loss, which is written as follows
\begin{subequations}\label{Eq:Guideline}
\begin{eqnarray}
&& \max ~ E = \P(T~\text{accepts}~H_0 ) \cdot R_{\text{privacy}}\label{Eq:guidelineefficiency}\\
&& \text{s.t.} ~~~ I_{\text{MLTS}} \cdot \P(\text{$T$ accepts $H_0$}) \leq \P(\text{RG succeeds}).\label{Eq:guidelinesecurity}
\end{eqnarray}
\end{subequations}

The design guideline is proposed to find the settings for test $T$ and the privacy amplification rate $R_{\text{privacy}}$ to maximize efficiency under the constraint of no security loss. \eqref{Eq:guidelinesecurity} ensures that $L_{\text{security}}=  0$ by selecting the proper P-value threshold $\alpha$ and $R_{\text{privacy}}$. We provide the theoretical analysis of how to calculate $\P(\text{$T$ accepts $H_0$})$ for different $\alpha$ values under different randomness tests in Appendix~C. Although we have both theoretical results of $I_{\text{MLTS}}$ and $\P(\text{$T$ accepts $H_0$})$, there is no straightforward convex or concave property (i.e., increasing $\alpha$ and decreasing $R_{\text{privacy}}$ may both satisfy \eqref{Eq:guidelinesecurity}). In practical systems, $\alpha$ and $R_{\text{privacy}}$ have typical value ranges and we select $\alpha \in [0.0001, 0.3]$ and $R_{\text{privacy}} \in [0.1, 1]$ in this paper. Within the ranges, we use greedy search with small granularity to find the best pair that maximizes \eqref{Eq:guidelineefficiency}. From the design guideline \eqref{Eq:Guideline}, we can answer the questions in Section~\ref{SubSec:TestingRandomness}.
\begin{enumerate}
\item The cryptographic domain function $f_c$ is based on privacy amplification, however, over-estimating the entropy cannot be avoided, since it is generally difficult to accurately estimate the per-bit entropy of a physical source \cite{turan2018recommendation}.  Consequently, if the randomness test $T$ is set in the cryptographic domain, it is equivalent to testing the output of a sufficiently random sequence, and always passing the test $T$.  Therefore, it is reasonable to test the wireless domain output $X = f_{w}(\omega_D )$ when extracting randomness from the wireless channel.
\item Based on \eqref{Eq:Guideline}, we can solve the optimization function to find the sufficient P-value threshold $\alpha$ and $R_{\text{privacy}}$, instead of manually setting the parameters, to {\hlk ensure no additional loss in security} and achieve high efficiency.
\end{enumerate}

Randomness test $T$ is an important part in the guideline~\eqref{Eq:Guideline}. Rather than designing a new randomness test, we focus on NIST randomness tests as they have been well structured and widely adopted \cite{mathur2008radio, jana2009effectiveness, wallace2010automatic, chen2011secret, liu2012collaborative, wang2011fast, wang2012cooperative, liu2013fast, tsouri2013threshold , zhu2013extracting, ali2014eliminating, 14pgk-seon, xi2016instant}. In order to configure the NIST randomness tests, we need to analyze the relationship between $\P(T~\text{accepts}~H_0 )$ and $(\rho, \alpha)$ for a specific test. Hence, in the following, we present how to bridge $\P(T~\text{accepts}~H_0 )$ to $h(\rho, \alpha)$, where $h(\cdot)$ represents the probability function of $(\rho, \alpha)$ for a specific NIST test.

In many scenarios, multiple randomness tests can be used together for testing. This combination can enhance security and indeed loosen the $R_{\text{privacy}}$ setting. However, the setup for a single test in existing studies is not loosened even when multiple tests are used. This has been common in existing studies for wireless key generation  \cite{mathur2008radio, jana2009effectiveness, wallace2010automatic, chen2011secret, liu2012collaborative, wang2011fast, wang2012cooperative, liu2013fast, tsouri2013threshold , zhu2013extracting, ali2014eliminating, 14pgk-seon, xi2016instant}, hardware security \cite{petrie2000noise}, cryptography and software security \cite{machicao2017improving, lacy1993cryptolib}. This is due to two major reasons: i) it can be mathematically intractable to analyze the joint correlation among multiple tests. The NIST guideline \cite{nisttest} performed such a correlation study and only shows that empirically the correction among NIST tests is very small. As a result, it can be difficult to show how much the setup for each test can be loosened analytically. ii) Using single-test setup can ensure the worst-case security guarantee even when multiple tests are used. We also adopt this practice in the paper and recommend the use of the single-test setup for multiple-test scenarios.

\subsection{Analysis of NIST Randomness Tests}\label{SubSec:Tests}
There are fifteen tests provided in the NIST test suite \cite{nisttest}, and we choose {\hlk nine} of them which are commonly used in the existing literature \cite{mathur2008radio, jana2009effectiveness,    wallace2010automatic, chen2011secret, liu2012collaborative, wang2011fast, wang2012cooperative, liu2013fast, tsouri2013threshold , zhu2013extracting, ali2014eliminating, 14pgk-seon, xi2016instant}. The nine tests in our study are frequency test, frequency test within a block, runs test, test for the longest run of ones in a block, discrete fourier transform test, non-overlapping template matching test, approximate entropy test and serial test (serial test has two orders). Based on the P-value computation formula, these tests can be categorized into two classes: Gaussian and chi-square distribution. In the following, we use the most common frequency test $T_{\text{freq}}$ as the representative to show the procedure of the relationship between $\P(\text{$T_{\text{freq}}$ accepts $H_0$})$ and $(\rho, \alpha)$. The results of other tests are provided in Appendix~\ref{Sec:TestResult}. We use the function $h_{\text{freq}}(\rho, \alpha)$ to represent frequency test in NIST. Given a bit sequence $X=[x_1, x_2, \dots x_L]$ from Definition~\ref{Def:BitGen},
\begin{eqnarray}\label{Eq:hfreq}
h_{\text{freq}}(\rho, \alpha) &=& \P(\text{$T_{\text{freq}}$ accepts $H_0$}) \nonumber\\
&=& \P(|S_{\text{freq}}(X)| < \sqrt{2} \text{erfc}^{-1}(\alpha)),
\end{eqnarray}
where $S_{\text{freq}}(X) = \frac{1}{\sqrt{L}}\sum_{l=1}^{L}(2x_l - 1)$ is the statistics definition of frequency test. Since the correlated sequence $X$ can be considered as generating from a uniformly ergodic Markov chain \cite{jones2004markov}, $S_{\text{freq}}(X)$ can be derived by the Markov central limit theorem, only if we know the mean and variance. $|S_{\text{freq}}(X)| \sim \mathcal{N}\left( 0, \frac{1+|\rho|}{1-|\rho|}\right)$ is followed by Gaussian distribution. Thus, we have the $h_{\text{freq}}(\rho, \alpha)$ as follows
\begin{eqnarray}
h_{\text{freq}}(\rho, \alpha) = \text{erf} \left( \text{erfc}^{-1}(\alpha) \sqrt{\frac{1-\rho}{1+\rho}} \right),
\end{eqnarray}
where $\text{erf}$ and $\text{erfc}^{-1}$ are error function and inverse complementary error function. Based on the analysis of randomness test $T$, we can choose the proper pairs of P-value threshold $\alpha$ and privacy amplification rate $R_{\text{privacy}}$.

\begin{figure*}[t!]
\centering
\begin{minipage}[c]{0.24\textwidth}
    \centering
    \includegraphics[width=\textwidth]{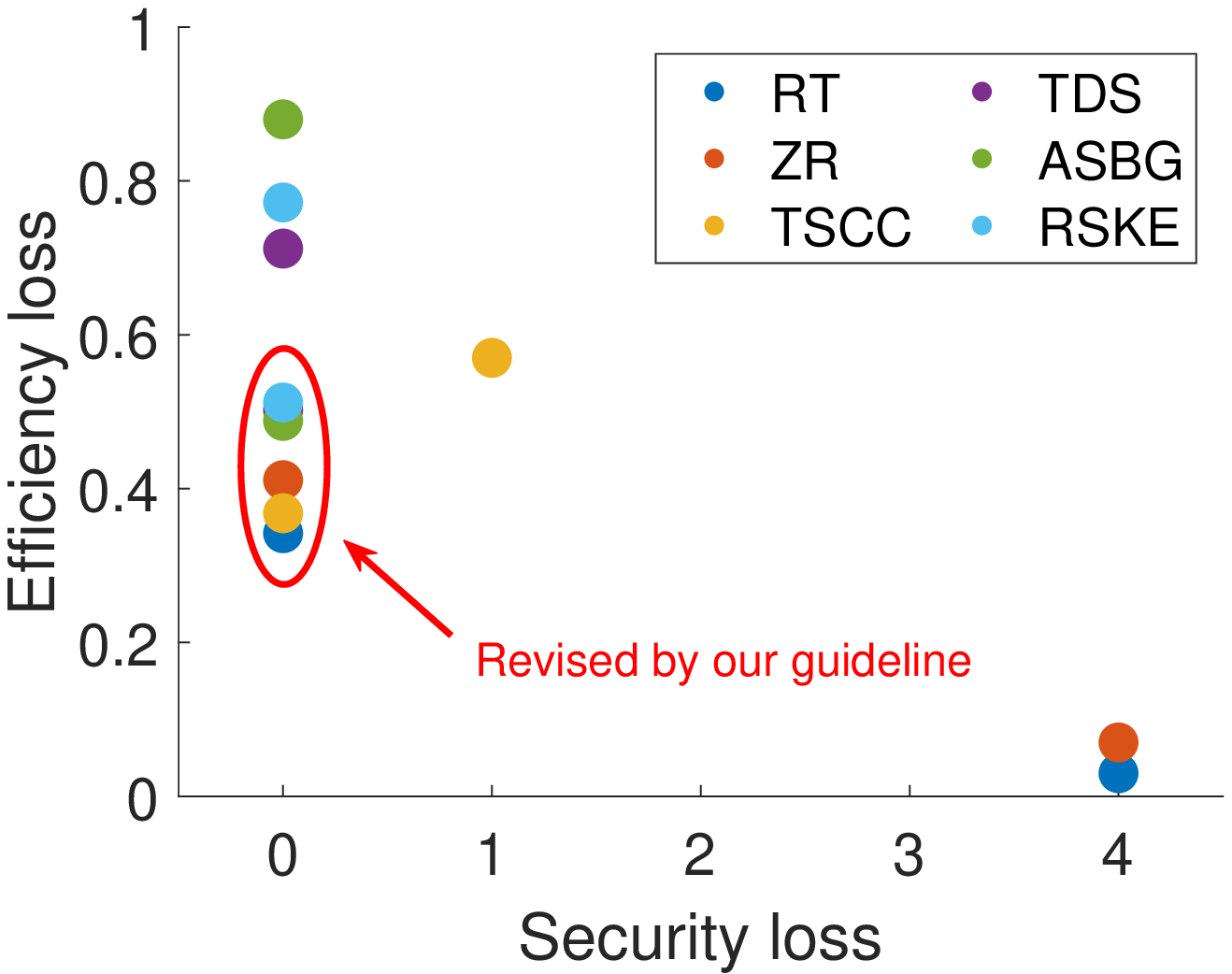}
     \caption{Security loss vs Efficiency loss}
     \label{Fig:SecurityRevised}
\end{minipage}
\hfill
\begin{minipage}[c]{0.24\textwidth}
    \centering
    \includegraphics[width=\textwidth]{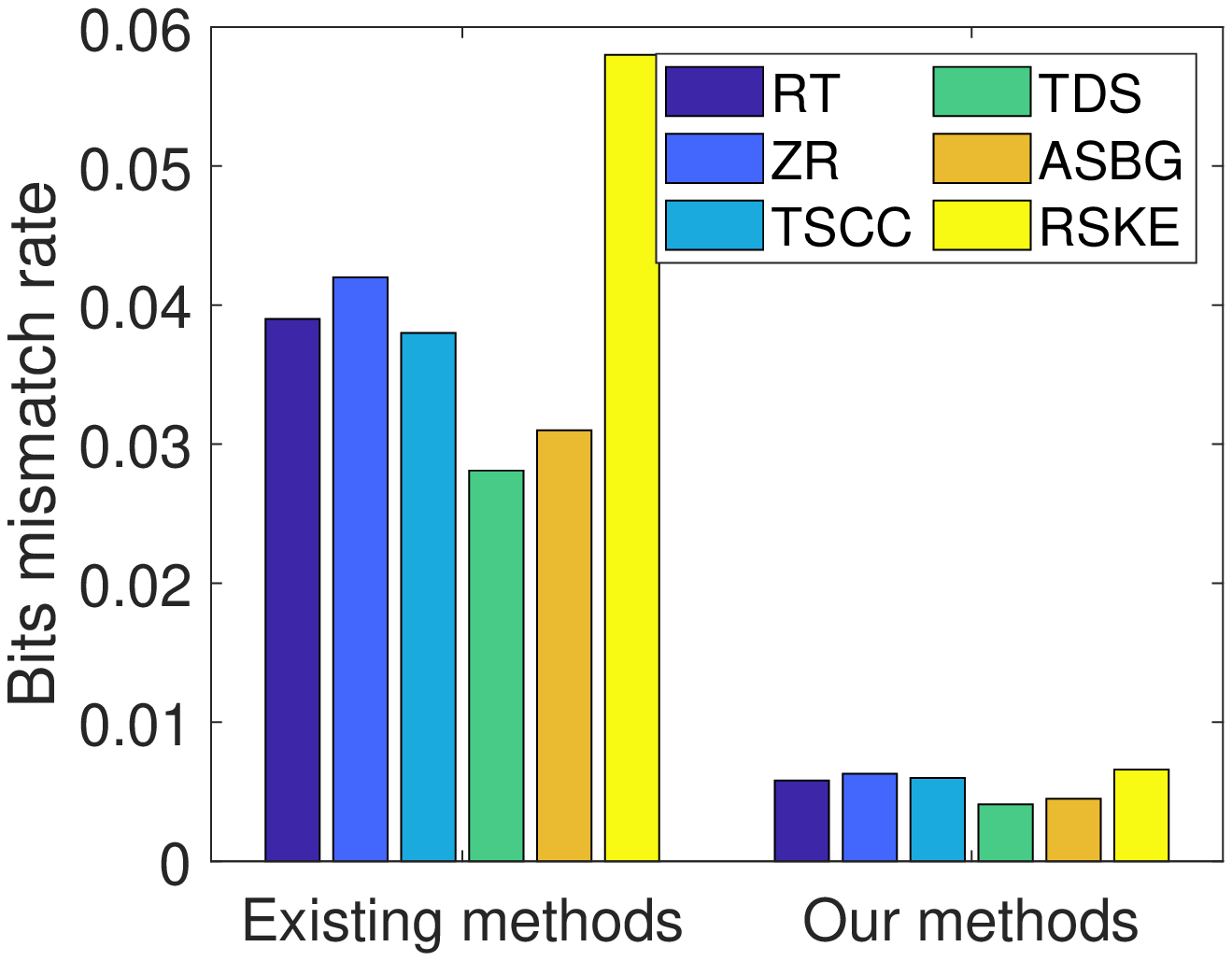}
     \caption{Bits mismatch rate}
     \label{Fig:MismatchRevised}
\end{minipage}
\hfill
\begin{minipage}[c]{0.24\textwidth}
    \centering
    \includegraphics[width=\textwidth]{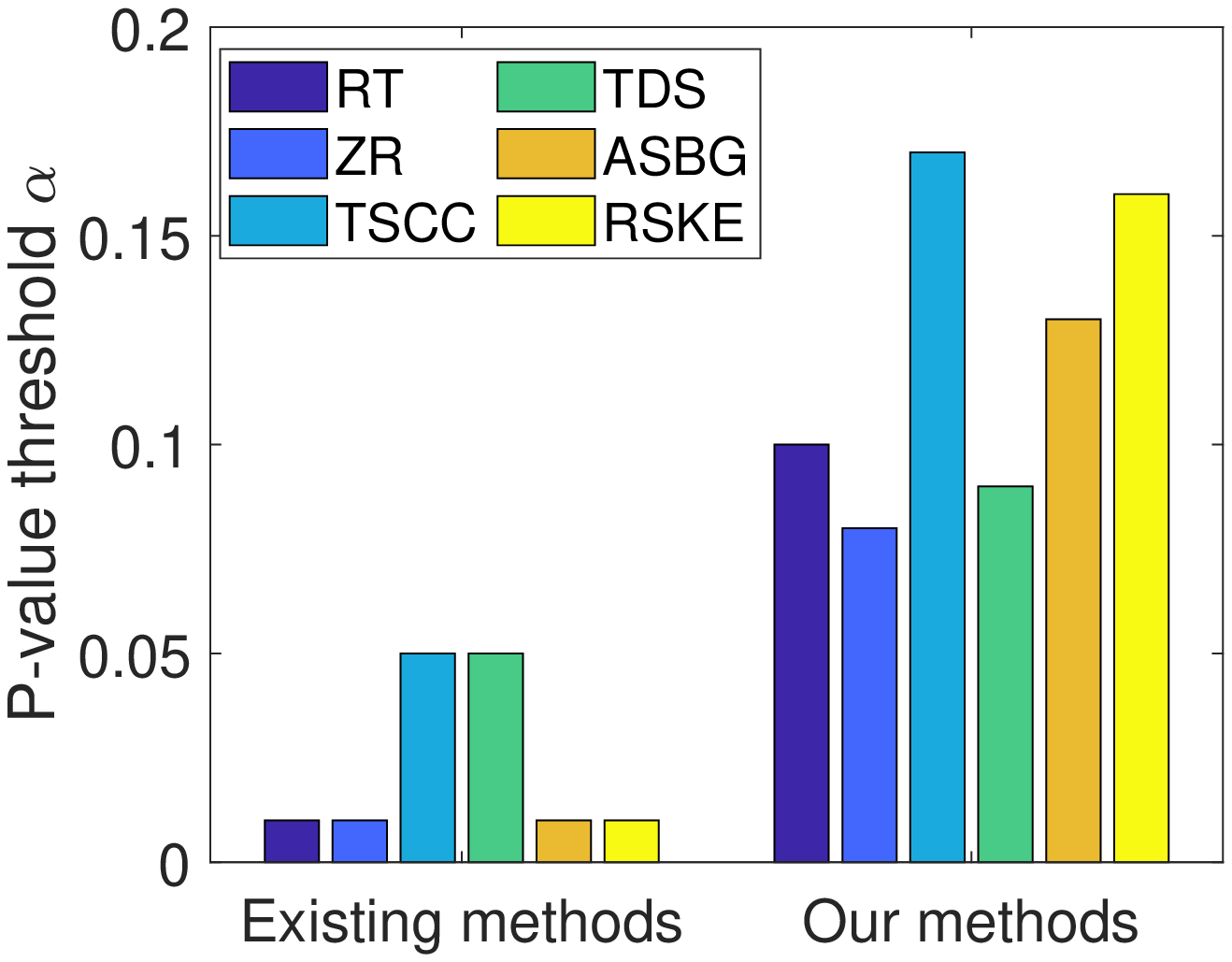}
     \caption{P-value threshold $\alpha$}
     \label{Fig:P-threshold}
\end{minipage}
\hfill
\begin{minipage}[c]{0.24\textwidth}
    \centering
    \includegraphics[width=\textwidth]{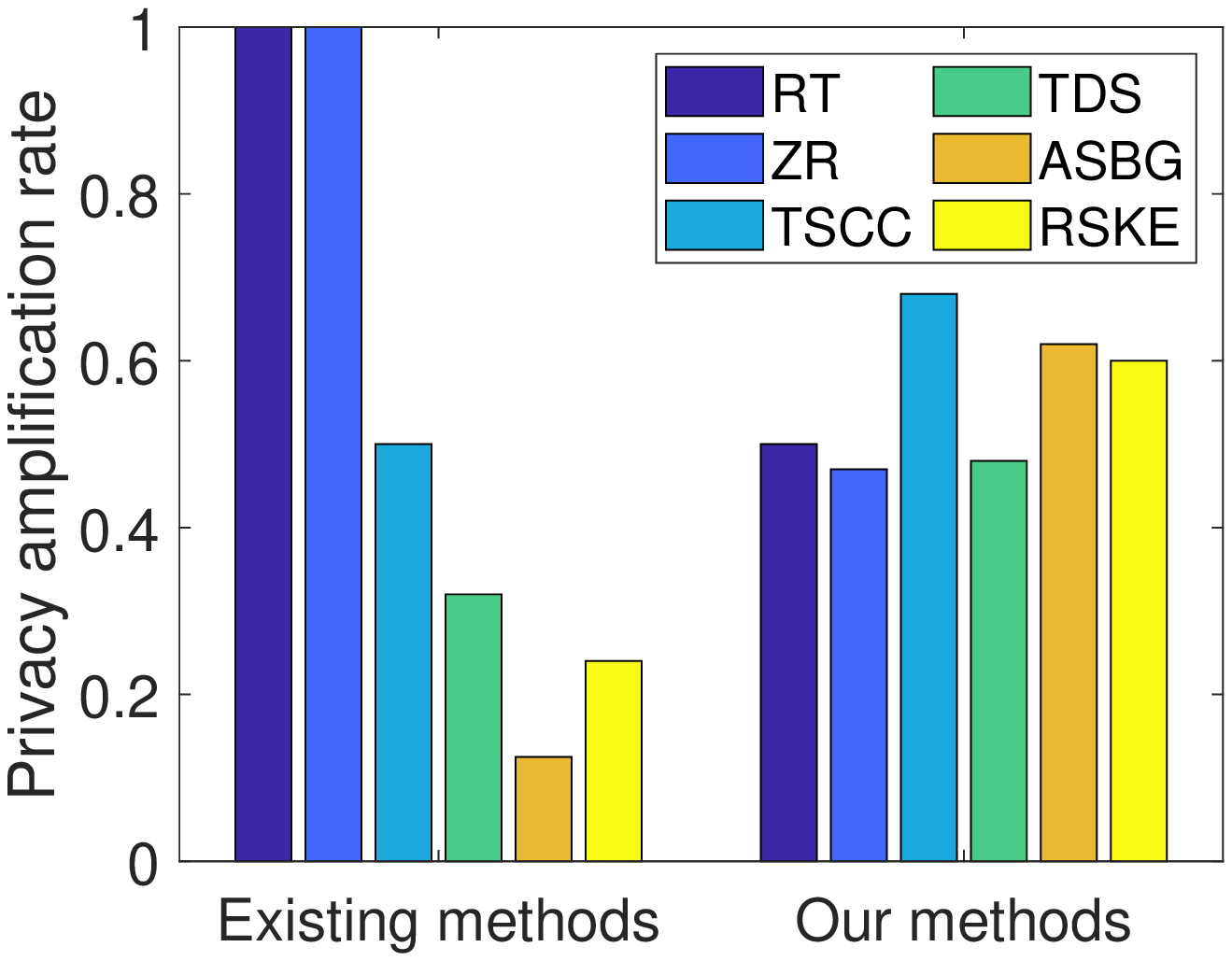}
     \caption{Privacy amplification rate $R_{\text{Privacy}}$}
     \label{Fig:PArate}
\end{minipage}
\end{figure*}

\section{Experimental Evaluation}\label{Sec:Exp}
In this section, we obtain the security loss, efficiency loss and bits mismatch rate of the secret key generation by the wireless channel response. In the following, we first introduce the experimental setup. Then, we compare the performance of existing secret key generation methods before and after being revised by our design guideline in \eqref{Eq:Guideline} under different experimental environments, different randomness tests and different length of keys.

\subsection{Experimental Setup}
\noindent\textbf{Channel Response Measurements:} The first step towards analyzing the randomness of the secret key generation in the wireless domain is to collect a large number of channel information (RSSI, CSI and Phase shifts) in realistic environments. We use two USRP X310s, acting as a transmitter and a receiver respectively, to build our experimental platform, where each device is equipped with a UBX-160 daughterboard and a VERT 2450 antenna. {\hlk The software toolkit is GNURadio. We implement a typical training data-aided time and frequency synchronization scheme based on \cite{schmidl1997robust} for channel probing whose procedure follows \cite{mathur2008radio}. For the equalization, we adopt a frequency-domain OFDM equalizer with the aid of pilot tones \cite{huang2007robust}.} To measure the channel information, the transmitter consistently sends training sequences (as known as the preamble in wireless standards) to the receiver with fixed transmit power. On this experiment platform, we collect more than 1 billion channel information in total spanning over 24 hours in different environments with 2.4GHz carrier frequency and 1.0MHz bandwidth.

\begin{table}[ht]
\centering
\caption{Parameters setting of existing methods.}\label{Tab:ExampleUse}
\resizebox{0.47\textwidth}{!}{\begin{tabular}{|c|c|c|c|c|}
\hline
Examples & Test setting domain & $\alpha$ value & $R_{\text{Privacy}}$ & Source\\
\hline
RT & wireless & 0.01 & 1 & RSSI\\
\hline
ZR & wireless & 0.01 & 1 & RSSI\\
\hline
TSCC & wireless & 0.05 & 0.5 & Phase\\
\hline
TDS & cryptographic & 0.05 & 0.32 & CSI\\
\hline
ASBG & cryptographic & 0.01 & 0.125 & RSSI\\
\hline
RSKE & cryptographic & 0.01 & 0.24 & RSSI\\
\hline
\end{tabular}}
\end{table}

\noindent\textbf{Secret Key Generation Model:} We compare the performance of 6 existing secret key generation models: Radio-telepathy (RT) \cite{mathur2008radio}, Zero Reconciliation (ZR) \cite{ali2014eliminating}, Temporally and Spatially Correlated Coefficients (TSCC) \cite{chen2011secret}, The Dancing Signals (TDS) \cite{xi2016instant}, Adaptive Secret Bit Generation (ASBG) \cite{jana2009effectiveness} and Robust Secret Key extraction (RSKE) \cite{14pgk-seon}, where the P-value threshold $\alpha$, $R_{\text{privacy}}$, different NIST statistical randomness test setting position and source of secret key generation are shown in Table~\ref{Tab:ExampleUse}.

\noindent\textbf{Eve's Capability:} The attacker Eve aims to obtain the secret key $K_D$ through MLTS without knowing any channel information. Although a realistically powerful capability for Eve is $2^{63.1}$ \cite{sha1attack2017}, it is still statistically insignificant in the experimentation in order to crack a long key, for example, the attack success probability of cracking a 128-bit key is about $2^{-64}$. Thus, we consider a more powerful capability of Eve with $N = 2^{96}$ such that attack success probability increases from $2^{-64}$ to $2^{-32}$, where is observable in our experiments.

\noindent\textbf{Evaluation Metrics:} The evaluation metrics used in our experiments are security loss $L_{\text{security}}$ and efficiency loss $L_{\text{efficiency}}$ defined in Section~\ref{SubSec:Keyestablish} as well as the bits mismatch rate $R_{\text{mismatch}}$, which is the ratio of the number of bits that do not match between Alice and Bob to the number of bits extracted from channel. All the experimental results are the average value from at least 30 independent experiments.

\subsection{Evaluation Results}
\noindent\textbf{Evaluation of Existing Secret Key Generation Methods:} We aim to show the performance (i.e., $L_{\text{security}}$ and $L_{\text{efficiency}}$ for 128-bit secret key) of existing methods with the P-value threshold $\alpha$ and $R_{\text{privacy}}$ settings in the literature in comparison with the new values for these parameters based on our design guideline. We collect the channel information under 5 meters laboratory environment and using frequency test.

\begin{figure*}[t!]
\centering
\begin{minipage}[c]{0.24\textwidth}
    \centering
    \includegraphics[width=\textwidth]{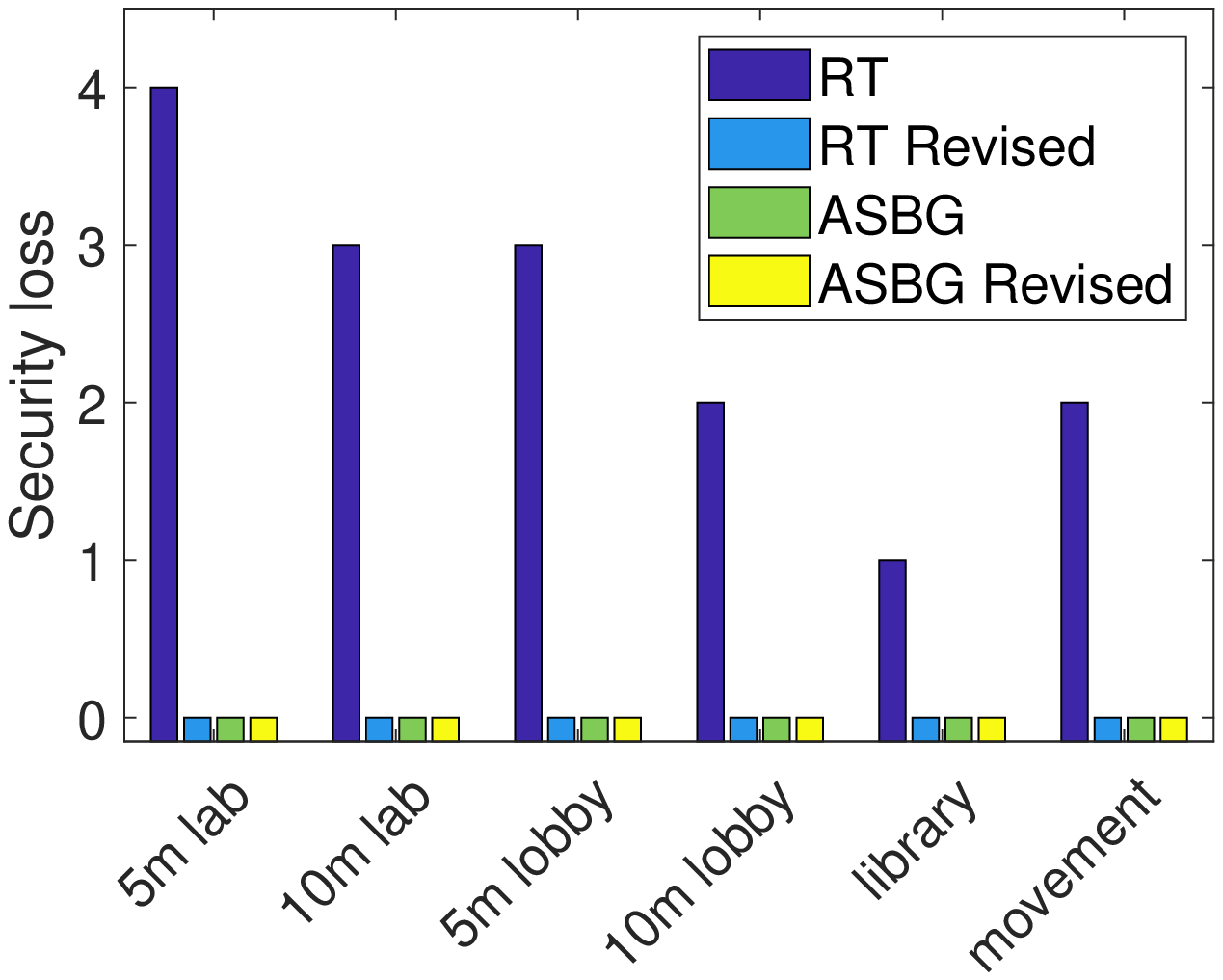}
     \caption{$L_{\text{security}}$ in different experimental environments.}
     \label{Fig:Security_vs_environment}
\end{minipage}
\hfill
\begin{minipage}[c]{0.24\textwidth}
    \centering
    \includegraphics[width=\textwidth]{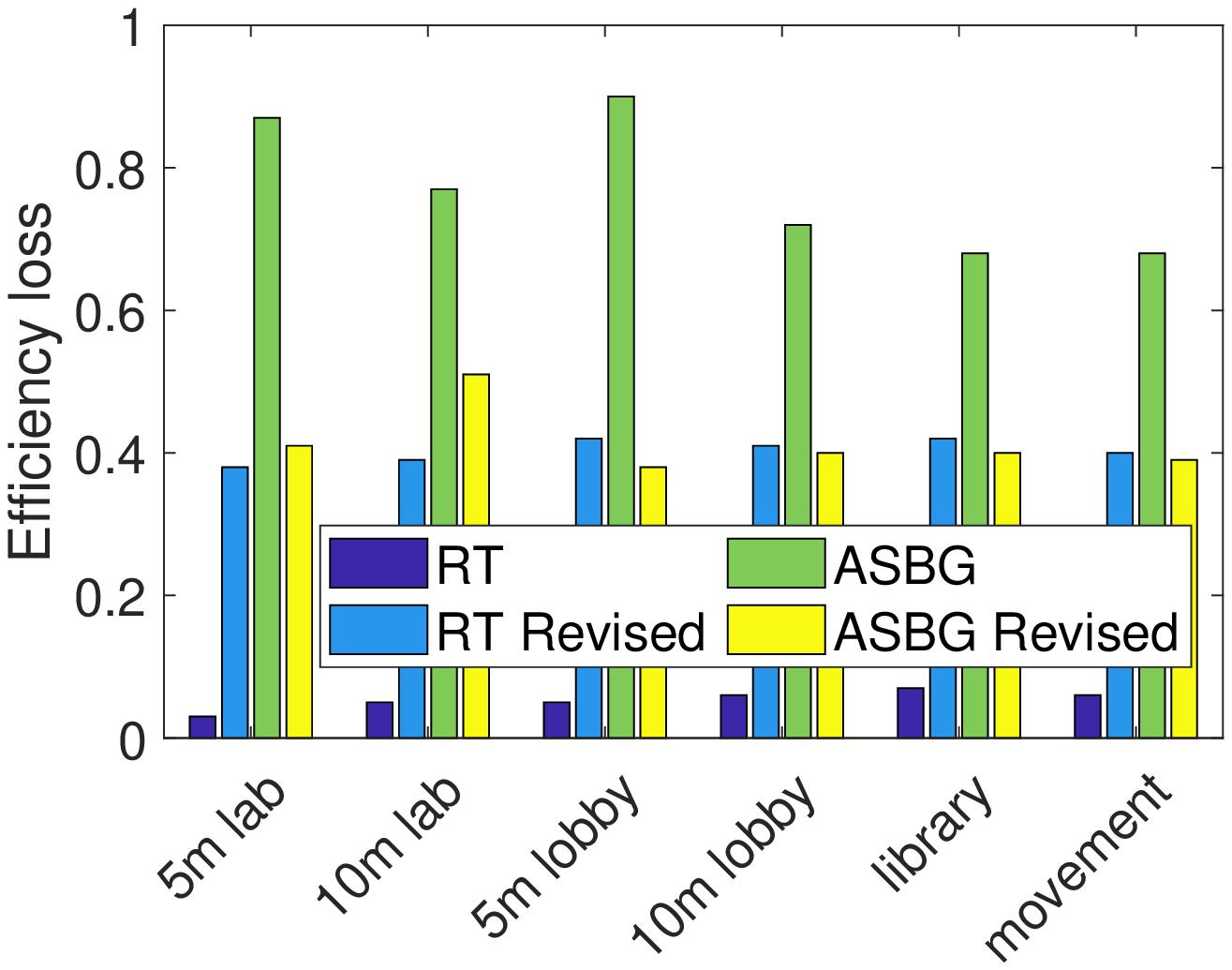}
     \caption{$L_{\text{efficiency}}$ in different experimental environments.}
     \label{Fig:Efficiency_vs_environment}
\end{minipage}
\hfill
\begin{minipage}[c]{0.24\textwidth}
    \centering
    \includegraphics[width=\textwidth]{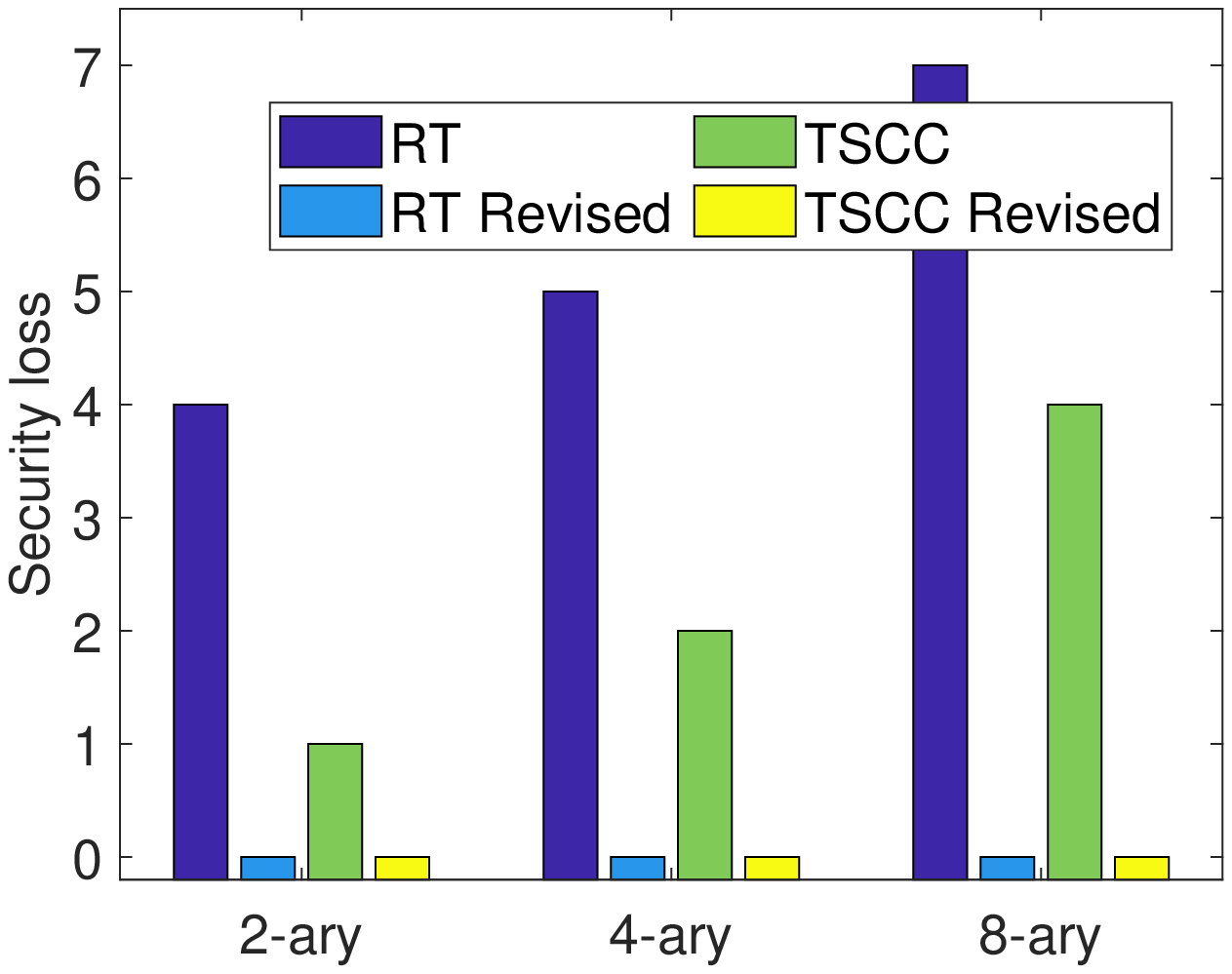}
     \caption{$L_{\text{security}}$ under multiple level quantizations.}
     \label{Fig:Security_vs_bit}
\end{minipage}
\hfill
\begin{minipage}[c]{0.24\textwidth}
    \centering
    \includegraphics[width=\textwidth]{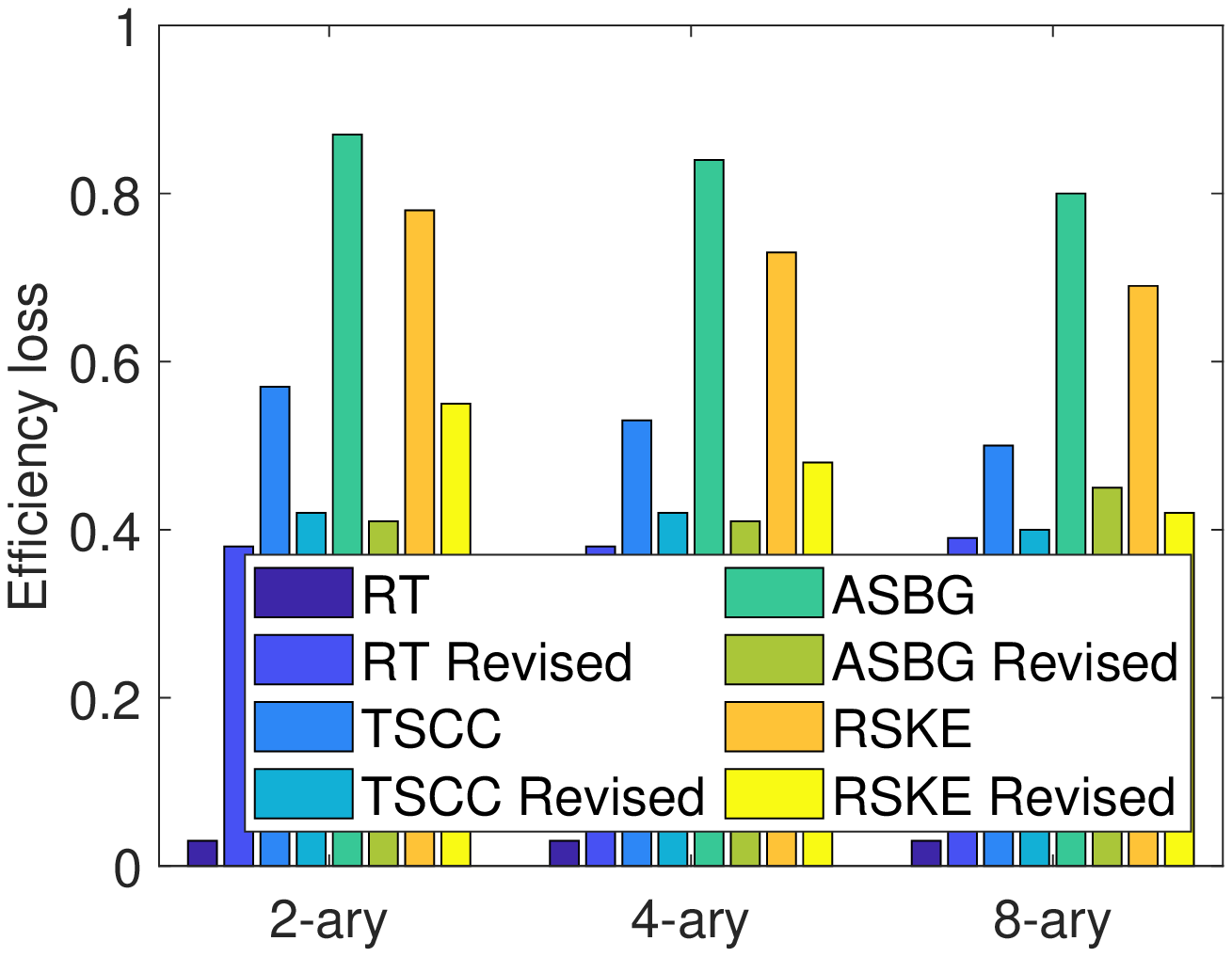}
     \caption{$L_{\text{efficiency}}$ under multiple level quantizations.}
     \label{Fig:Efficiency_vs_bit}
\end{minipage}
\end{figure*}

\begin{figure*}[t!]
\centering
\begin{minipage}[c]{0.32\textwidth}
    \centering
    \includegraphics[width=\textwidth]{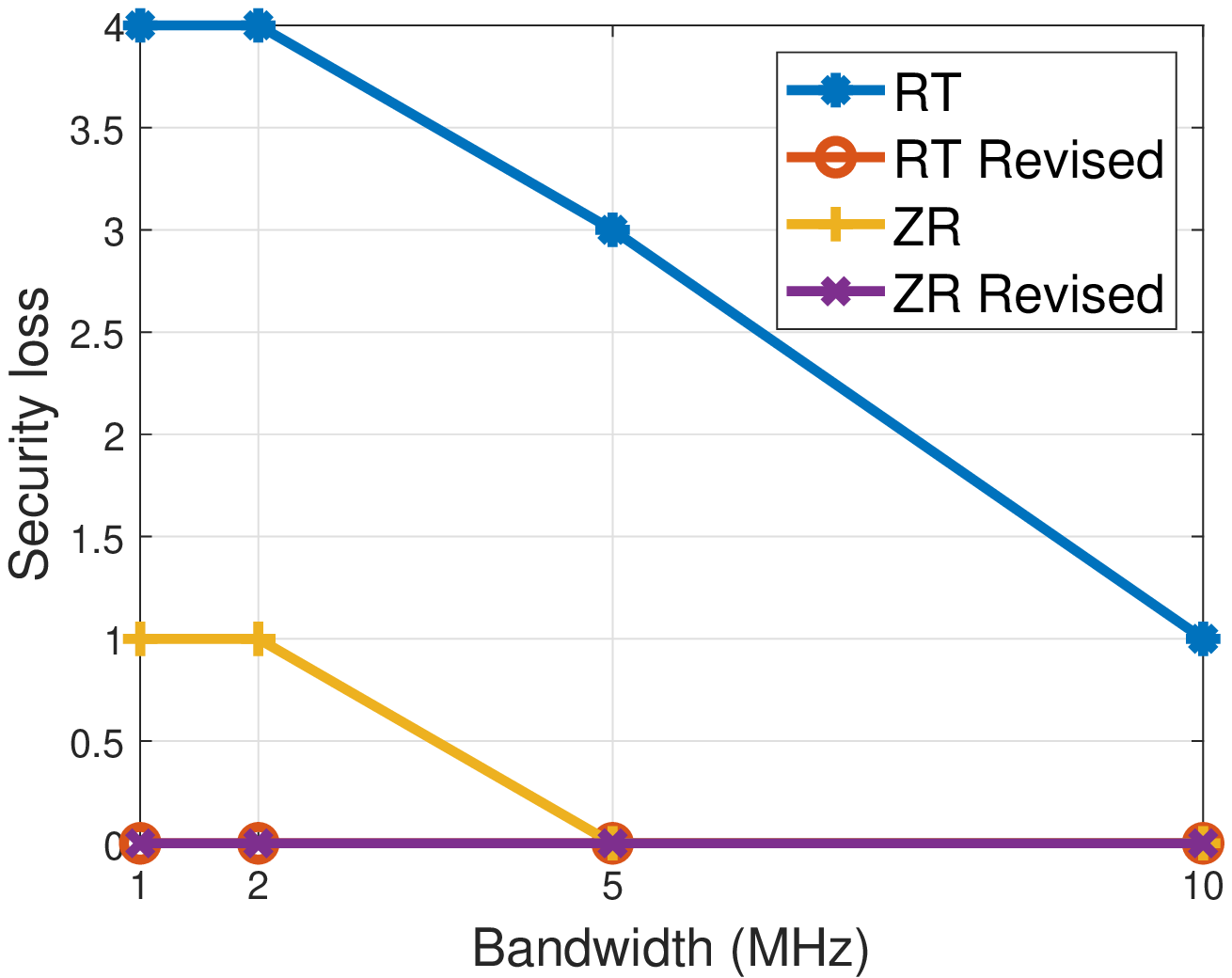}
     \caption{$L_{\text{security}}$ under different bandwidths.}
     \label{Fig:bandwidth_security}
\end{minipage}
\hfill
\begin{minipage}[c]{0.32\textwidth}
    \centering
    \includegraphics[width=\textwidth]{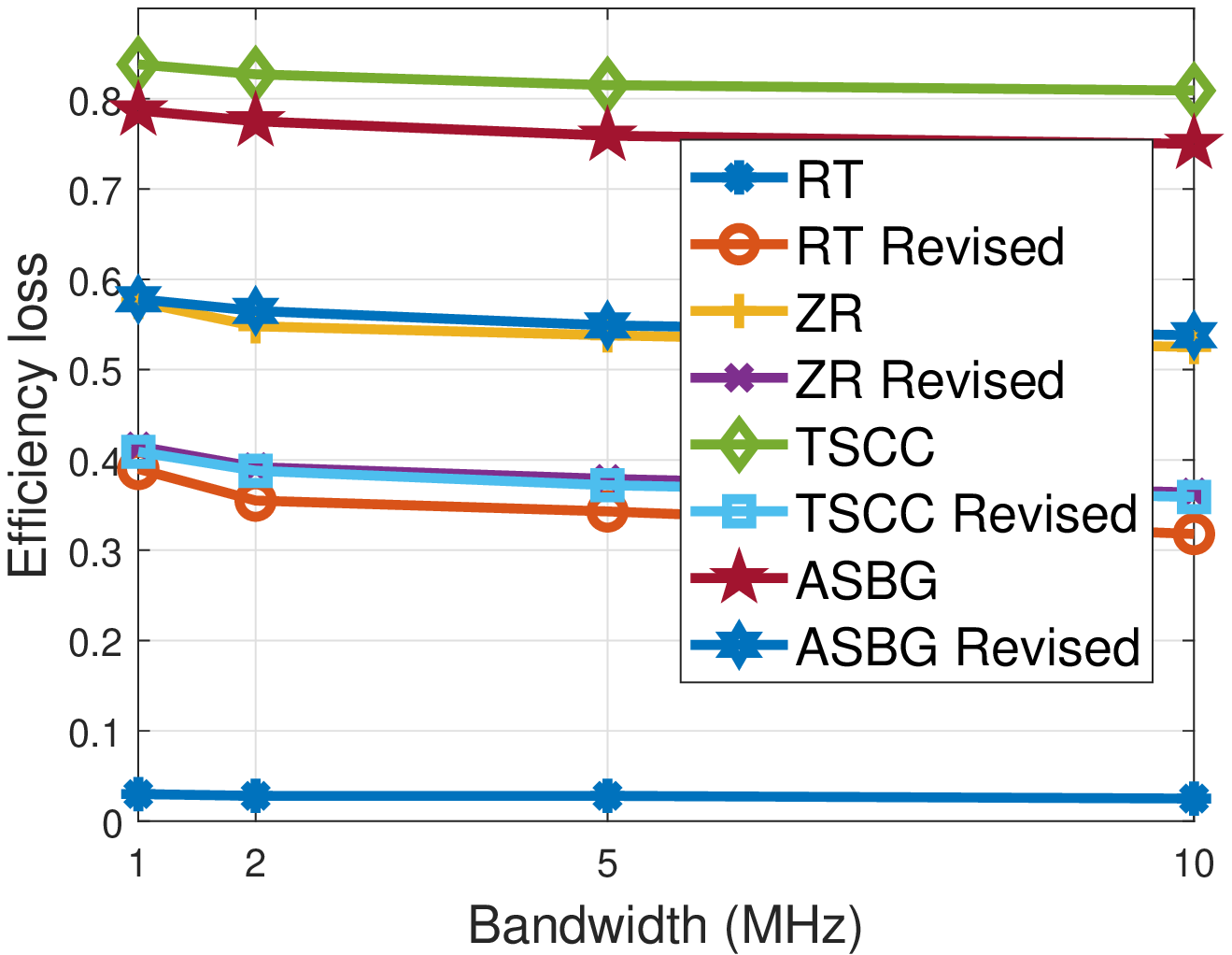}
     \caption{$L_{\text{efficiency}}$ under different bandwidths.}
     \label{Fig:bandwidth_efficiency}
\end{minipage}
\hfill
\begin{minipage}[c]{0.32\textwidth}
    \centering
    \includegraphics[width=\textwidth]{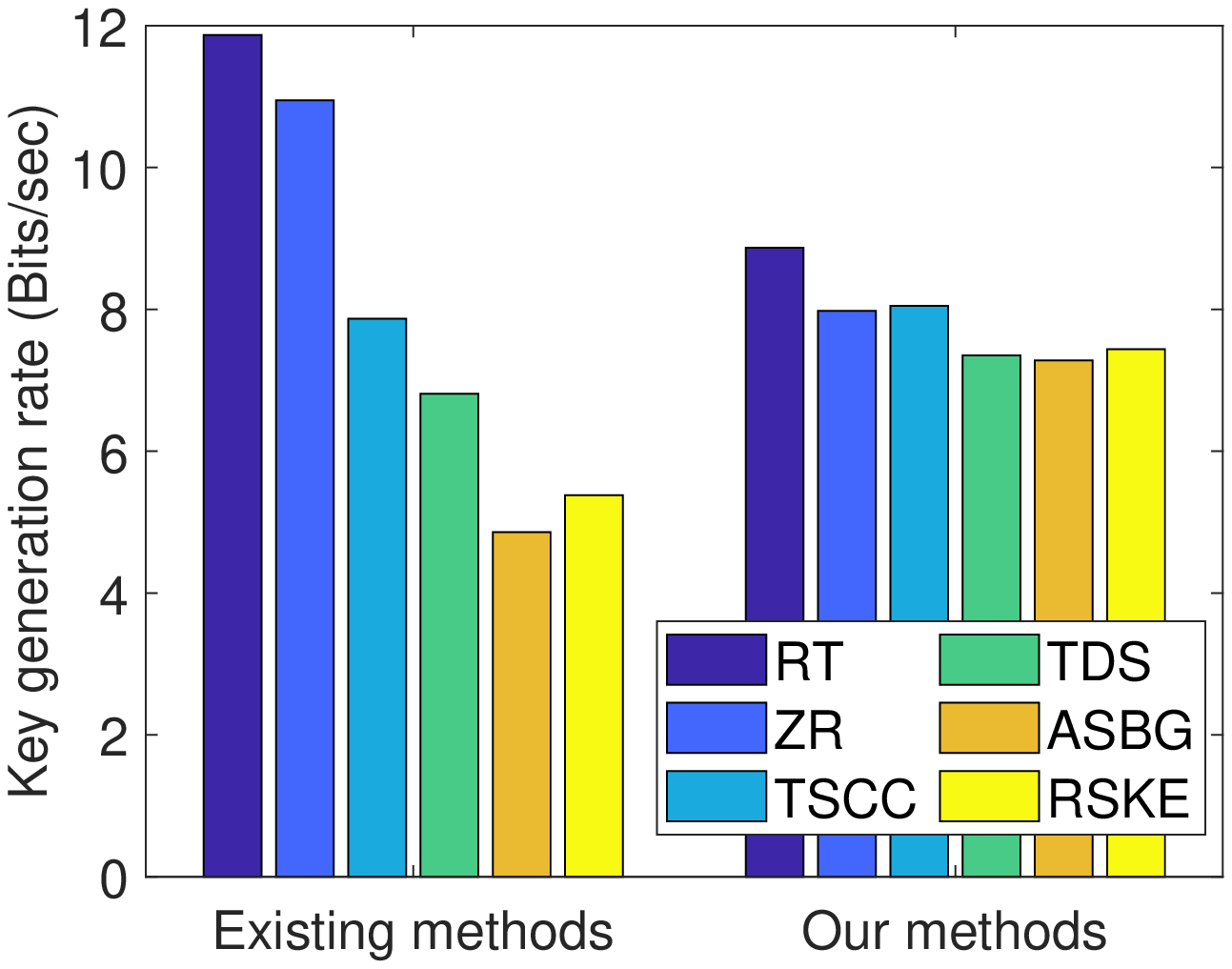}
     \caption{Key generation rate (Bit/sec) of different methods.}
     \label{Fig:rate}
\end{minipage}
\end{figure*}

Figure \ref{Fig:SecurityRevised} shows that the security and efficiency losses of these seven different secret key generation methods before and after new parameter settings based on our guideline. We can see that RT, ZR and TSCC, which set the tests in wireless domain, have  higher $L_{\text{security}}$ than others in the cryptographic domain, e.g., $L_{\text{security}}$ of RT is $4$. Although setting test $T$ in the cryptographic domain can eliminate security loss, it leads to high efficiency loss, e.g, ASBG has $L_{\text{efficiency}} = 0.883$. After the new settings based on our design guideline, i) all of these methods have no $L_{\text{security}}$; ii) $L_{\text{efficiency}}$ significantly decreases. For example, ASBG changes from $0.883$ to $0.487$ in terms of efficiency loss. Through the revision of parameters based on our guideline, $R_{\text{mismatch}}$ of all methods can be reduced, as shown in Figure~\ref{Fig:MismatchRevised}. Figures~\ref{Fig:P-threshold} and \ref{Fig:PArate} show the P-value threshold $\alpha$ and $R_{\text{privacy}}$ are calculated by our design guideline, which indicate that different secret key generation methods need to calculate different $\alpha$ and $R_{\text{privacy}}$ to eliminate security loss and achieve high efficiency. On the other hand, Figures~\ref{Fig:P-threshold} and~\ref{Fig:PArate} also illustrate that if we choose a higher P-value threshold, we also need a higher $R_{\text{privacy}}$ to eliminate security loss.

\noindent\textbf{Evaluation of Different Experimental Environments:} The secret key generation method may extract the secret keys with different randomness degrees under varying wireless communication environments. We will show $L_{\text{security}}$ and $L_{\text{efficiency}}$ with or without our design guideline in different practical scenarios. Among the 6 secret key generation methods, we select RT and ASBG, since they set the NIST test at the different positions. We conduct experiments to collect RSSI, CSI or phase shift in different indoor environments. For the static USRPs setting, we set the distance of two USRPs to be 5 and 10 meters in 40m$^2$ laboratory, 10 and 25 meters in 350m$^2$ public lobby, 25 meters in 450m$^2$ campus library. For the movement setting, we fix the position of one USRP, and randomly move another one in laboratory. We consider the laboratory environment line-of-sight condition, and lobby and library environments non-line-of-sight condition.

Figure~\ref{Fig:Security_vs_environment} shows $L_{\text{security}}$ under different experimental environments, and that RT performs the worst for each environment. After revision by our guideline, we eliminate all security loss (i.e., $L_{\text{security}} = 0$). From Figure~\ref{Fig:Efficiency_vs_environment}, we increase the $L_{\text{efficiency}}$ from $0.05$ to around $0.4$ of RT under each communication environment. In ASBG, since they set a small original $R_{\text{privacy}}$ value (i.e., $0.125$ in Table~\ref{Tab:ExampleUse}), it incurs high $L_{\text{efficiency}} = \{0.85, 0.74, 0.87, 0.71, 0.68, 0.69\}$ under different environments. After revision by our guideline, we properly set the new $R_{\text{privacy}}$ so that ASBG performs more efficiently such that $L_{\text{efficiency}} = \{0.41, 0.49, 0.37, 0.39, 0.39, 0.38\}$.

\noindent\textbf{Evaluation of Multiple Level Quantizations:} Figures~\ref{Fig:Security_vs_bit} and \ref{Fig:Efficiency_vs_bit} show $L_{\text{security}}$ and $L_{\text{efficiency}}$ under $m$-ary quantization methods $(m \in \{2, 4, 8\})$ with 4 methods: RT, TSCC, ASBG and RSKE.


Due to the fact that ASBG and RSKE with original settings have no security loss, we do not show them in Figure~\ref{Fig:Security_vs_bit} for clearer illustration. In Figure~\ref{Fig:Security_vs_bit}, we can see that a larger value of $m$ incurs a higher security loss (e.g., for RT, $L_{\text{security}} = 7$ when $m=8$). After revision based on our guideline, we ensure $L_{\text{security}}=0$ for RT and TSCC. In Figure~\ref{Fig:Efficiency_vs_bit}, it is observed that when we ensure $L_{\text{security}}=0$ by setting up randomness tests for different values of $m$, the efficiency $L_{\text{security}}$ approximately remains the same. This indicates that simply adopting a multiple level quantization method does not necessarily mean a faster key generation rate with security guarantee.

\begin{table*}[t!]
\centering
\caption{P-value threshold $\alpha$ and $R_{\text{privacy}}$ setting for different quantization methods.}
\resizebox{0.89\textwidth}{!}{\begin{tabular}{|c|c|c|c|c|c|c|c|c|c|c|c|c|}
\hline
                    & \multicolumn{2}{c|}{RT} & \multicolumn{2}{c|}{ZR} & \multicolumn{2}{c|}{TSCC} & \multicolumn{2}{c|}{TDS} & \multicolumn{2}{c|}{ASBG} & \multicolumn{2}{c|}{RSKE} \\ \hline
                    & \hlk{$\alpha$}    & \hlk{$R_{\text{privacy}}$}      & \hlk{$\alpha$}  & \hlk{$R_{\text{privacy}}$} & \hlk{$\alpha$}   & \hlk{$R_{\text{privacy}}$}  & \hlk{$\alpha$}      & \hlk{$R_{\text{privacy}}$}  & \hlk{$\alpha$}      & \hlk{$R_{\text{privacy}}$}  & \hlk{$\alpha$}      & \hlk{$R_{\text{privacy}}$} \\ \hline
Frequency           & 0.11       & 0.50        & 0.08       & 0.47       & 0.17        & 0.68        & 0.09        & 0.48       & 0.13        & 0.62        & 0.16        & 0.60        \\ \hline
Block of Frequency  & 0.21       & 0.67       & 0.18       & 0.59       & 0.27        & 0.79        & 0.19        & 0.61       & 0.25        & 0.72        & 0.20        & 0.63        \\ \hline
Run                 & 0.14       & 0.51       & 0.09       & 0.51       & 0.22        & 0.70        & 0.11        & 0.52       & 0.14        & 0.63        & 0.14        & 0.57        \\ \hline
Longest Run of ones & 0.13       & 0.48       & 0.18        & 0.57       & 0.21        & 0.66        & 0.10        & 0.47       & 0.17        & 0.61        & 0.17        & 0.66        \\ \hline
DFT                 & 0.07       & 0.61       & 0.14       & 0.62       & 0.14        & 0.71        & 0.09        & 0.54       & 0.13        & 0.62        & 0.11        & 0.68        \\ \hline
Non-overlap         & 0.08       & 0.46       & 0.20       & 0.55       & 0.17        & 0.61        & 0.17        & 0.50       & 0.22        & 0.66        & 0.21        & 0.63        \\ \hline
Approximate entropy & 0.06       & 0.41       & 0.18       & 0.53       & 0.14        & 0.62        & 0.15        & 0.51       & 0.19        & 0.62        & 0.15        & 0.61        \\ \hline
First order serial  & 0.06       & 0.44       & 0.11       & 0.41       & 0.15        & 0.61        & 0.09        & 0.44       & 0.11        & 0.57        & 0.08        & 0.59        \\ \hline
Second order serial & 0.04       & 0.40       & 0.09       & 0.42       & 0.14        & 0.60        & 0.11        & 0.47       & 0.08        & 0.51        & 0.09        & 0.61        \\ \hline
\end{tabular}}
\end{table*}

\noindent\textbf{Evaluation of Different Bandwidths:} Figures~\ref{Fig:bandwidth_security} and \ref{Fig:bandwidth_efficiency} show $L_{\text{security}}$ and $L_{\text{efficiency}}$ under different bandwidth settings $\{1, 2, 5, 10\}$MHz in the lobby environment. We do not show the $L_{\text{security}}$ of ASBG and RSKE, since they have no security loss. When we increase the bandwidth, it is easy to observe that $L_{\text{security}}$ decreases under RT and TSCC (e.g., for RT, $L_{\text{security}}$ goes from 4 to 1). We can observe that the wireless signal with a larger bandwidth make the random bit sequence less correlated and therefore more random. From the efficiency perspective, increasing bandwidth can decrease $\P(\text{$T$ accepts $H_0$})$ such that $L_{\text{efficiency}}$ becomes smaller (e.g., for ASBG, $L_{\text{efficiency}}$ decreases from 0.57 to 0.52). 

\noindent\textbf{Evaluation of Key Generation Rates:} For testing the key generation rate of each method, we setup the 6 existing key generation methods for extracting 128-bit key sequence using the parameters in Table~\ref{Tab:ExampleUse} under 5 meters laboratory environment. Under the same environment, we revise the parameters by our guideline and test the key generation rate. For each experiment, we collect 10,000 bits and calculate the rate (bits/sec). The results are shown in Figure~\ref{Fig:rate}. We note that RT and ZR has higher rates with their original setups but they both have security losses. Our setups ensure the maximum key generation rates (e.g., 7.49 bits/sec for ASBG) without security loss.

\begin{table}[t!]
\centering
\caption{\hlk{$L_{\text{efficiency}}$ of different bit sequence length before and revised by our guideline.}}
\resizebox{0.48\textwidth}{!}{\begin{tabular}{|c|c|c|c|c|c|c|}
\hline
     & \multicolumn{2}{c|}{128-bit} & \multicolumn{2}{c|}{256-bit} & \multicolumn{2}{c|}{512-bit} \\ \hline
\hlk{$L_{\text{efficiency}}$} & \hlk{Original}  &  \hlk{Revised}  & \hlk{Original}  &  \hlk{Revised} & \hlk{Original}  &  \hlk{Revised}  \\ \hline

RT   & 0.06          & 0.33         & 0.03          & 0.35         & 0.08          & 0.31         \\ \hline
ZR   & 0.09          & 0.39         & 0.07          & 0.33         & 0.05          & 0.25         \\ \hline
TSCC & 0.58          & 0.37         & 0.57          & 0.31         & 0.55          & 0.24         \\ \hline
TDS  & 0.71          & 0.57         & 0.71          & 0.48         & 0.73          & 0.30         \\ \hline
ASBG & 0.86          & 0.42         & 0.88          & 0.36         & 0.85          & 0.29         \\ \hline
RSKE & 0.73          & 0.57         & 0.70          & 0.44         & 0.68          & 0.21         \\ \hline
\end{tabular}}
\end{table}

\noindent\textbf{Evaluation of Different Randomness Tests $T$ and $R_{\text{privacy}}$:} Table~II indicates that no tests can achieve the optimization goal in \eqref{Eq:Guideline} by using $\alpha = 0.01$. In other words, if we just follow the recommendation of NIST, we may not {\hlk eliminate the security loss and guarantee} efficiency of secret key generation. On the other hand, if we only focus on the security, the efficiency can be very low. Therefore, it is necessary to meet our guideline to consider both efficiency and security aspects. It is noted that Table~II shows different values of $\alpha$ and $R_{\text{privacy}}$ for different randomness tests. This is due to the fact that each  test has its own $\P(T~\text{accept}~H_0)$, which is provided in Appendix~\ref{Sec:TestResult}, in our guideline and we need to solve its own pair of $\alpha$ and $R_{\text{privacy}}$ based on the optimization \eqref{Eq:guidelineefficiency}.

\noindent\textbf{Efficiency of Different Length of Secret Keys Generation:} Intuitively, if Eve's attack capability does not change, it should be more difficult to crack a longer key sequence. Thus, randomness test $T$ and privacy amplification should be set looser for a longer key sequence. However, the existing methods use the fixed value $\alpha$ and $R_{\text{privacy}}$ for generating different lengths of key sequence such that it is detrimental to the secret key generation efficiency (suppose no security loss). We evaluate $L_{\text{efficiency}}$ of 128, 256 and 512-bit secret key generation before and after revision based on our guideline under the laboratory scenario and frequency test. Eve's attack capability is $N = 2^{96}$.

In Table~III, we can observe that it is more efficient to generate a longer secret key sequence based on our guideline (e.g., for TDS with revised setups, $L_{\text{efficiency}}$ is 0.57, 0.48 and 0.30 for the 128-bit, 256-bit and 512-bit cases, respectively). However, if we use the parameter setting in the existing studies, $L_{\text{efficiency}}$ does not change obviously, e.g., $L_{\text{efficiency}}$ of RSKE are $0.73$, $0.70$ and $0.68$. Hence, our design guideline also offers an adaptive method to generate keys with different lengths, which is not presented in existing studies.

\section{Related Work}\label{Sec:RW}

To {\hlk provide} the confidentiality of data transmission, secret key generation based on the information of wireless channels is promising because of the efficiency and security \cite{taha2015physical, premnath2014efficient, zhang2016efficient, luo2016rss}. In \cite{taha2015physical}, proximity attack requires the minimal distance from the eavesdropper to maintain perfect secrecy for secret key generation. The randomness test can provide a generic threshold on required distances from an eavesdropper and good key refreshing rates. \cite{premnath2014efficient} explores the use of wireless channel characteristics for establishing arbitrary length secret keys between Bluetooth devices. They verified the output secret bit streams generated by Bluetooth achieve high entropy by the randomness test. \cite{zhang2016efficient} tests the randomness of key bits, which quantifies a subcarrier's channel response with different coherent time. \cite{luo2016rss} proposes to defend against threats of eavesdropping and fake data injection in underwater acoustic networks, providing an overview of the advantages of RSSI based key generation and exploring the major challenges from the unique features of acoustic communications.

Key establishment using physical layer characteristics \cite{wang2015survey, wilson2007channel, huang2015dynamic}, which are much richer source of secret information but high computational cost overhead. \cite{wang2015survey} reviews different types of existing methods based on quantization, handling communication errors and the feasibility and security issues related to these methods. \cite{wilson2007channel} proposes the reciprocity theorem, which has become the most important theorem in this kind of method. The paper \cite{huang2015dynamic} provides an efficient secret key generation method using multipath relative delay from Ultra-wideband (UWB) channels. They study a statistical characterization of UWB channels in a residential scenario, and evaluate key mismatch probability. \cite{li2017secret} presents the key establishment that uses the distance variation trends caused by the motion paths of two devices to each other.

Recently, some studies have started working in authenticating the transmitter and receiver based on prior coordination or secret sharing. \cite{paul2008physical} proposes physical-layer authentication schemes through adding low-power signal. \cite{zhao2020faster} solves the authentication in IoT by exploiting the fading of the wireless links between devices to be authenticated and a set of trusted anchor nodes. \cite{argyraki2013creating} proposes a retroactive key setup to protect source authentication and path validation into the realm of practicality. \cite{paul2015wireless, liu2016physical} adapt fingerprint embedding to keep message authentication and increased security by obscuring the authentication tag.

\section{Conclusion}\label{Sec:Con}
This paper studies how to properly test the wireless channel randomness for security and efficiency. In particular, we proposes a new design guideline that can choose the P-value threshold, a critical parameter of the randomness test, to ensure the security of the wireless system as well as achieve a high secret bit generation rate with privacy amplification. Since the practical channel information (CSI, RSSI or phase shifts) is imperfectly independent, we come up with a new cracking key attack called MLTS, which searches the bit sequence by leveraging the Markov dependent property. By tuning a suitable P-value threshold and privacy amplification rate, we formulate an optimization problem to maximize the key generation rate under the constraint of no security loss. Our analysis indicates that the randomness test $T$ should be set in the wireless domain. We conduct different practical environments to validate the analysis of our guideline. By comparing to existing key generation methods, results show that our guideline can improve these methods to be more efficient and secure.


%

\appendices
\section{Proof of Theorem~\ref{TM:MLTS}}\label{Sec:ProofTheorem}
Because we cannot know the correlation coefficient exactly, we need to consider the positive and negative correlation simultaneously. Let $\theta$ be the transition probability on the generation tree, and we assume $\theta < 0.5$. In \cite{lindqvist1978note}, they proved that $\theta = \frac{1 - \rho}{2}$. Because Eve cannot know the bit sequence is positive correlation or not, she needs to search from the most likely bit sequences happen with probabilities $\theta^L$ and $(1 - \theta)^L$ simultaneously. MLTS searches for $X$ compute $K_T=f_c(X)$ from the most likely bit sequence towards the least likely one in $\{0,1\}^L$. Given the fact $K_D$ has been established, the MLTS success probability searching for $K_D$ is
\begin{eqnarray}
&& \!\!\!\!\!\!\!\!\! \P(\text{MLTS succeeds} ~|~ \text{$K_D$ established}) \nonumber\\
&& = \sum_{i=0}^{n/2} {L \choose i} \theta^i (1-\theta)^{L-i} + \sum_{i=0}^{n/2} {L \choose i} (1 - \theta)^i \theta^{L-i} \nonumber\\
&& = I_{\frac{1-\rho}{2}}\left(L-\frac{n}{2}, \frac{n}{2}+1\right) + I_{\frac{1+\rho}{2}}\left(L-\frac{n}{2}, \frac{n}{2}+1\right),
\end{eqnarray}
where $0 \leq n \leq L$, and $I_x(a,b)$ is the regularized incomplete beta function that has been defined in \eqref{Eq:U_xab}. \hfill~$\Box$

\section{MLTS for Multi-level Quantization}\label{Sec:multi}
If we use $m$ levels to quantify the wireless information, each level can be represented by a $b$-bit string, where $m = \log_2 b$. The multi-level quantization is also called $m$-ary quantization. Here, we redefine the correlation coefficient $\rho_m$ of consecutive bit arrays $x_i^m$ and $x_{i+1}^m$, where $x_i^m = \{x_{i,1}^m ,\dots, x_{i,m}^m \}$ in bit sequence $X$ as follows
\begin{equation}
    \rho_{m} = \frac{\sum_{j=1}^{m}(x_{i,j}^m - \bar{x}_i^m )(x_{i+1,j}^m - \bar{x}_{i+1}^m )}{\sqrt{\sum_{j=1}^{m}(x_{i,j}^m - \bar{x}_i^m )^2}\sqrt{\sum_{j=1}^{m}(x_{i+1,j}^m - \bar{x}_{i+1}^m )^2}}.
\end{equation}
For the $b$-ary quantization, transition probability $\theta_{r,s}$ (i.e., $\P(x_{i+1}^m = s | x_{i}^m = r)$), where $r$ and $s$ are the states, $r,s \in \{1,2,\dots,m\}$ in transition matrix $\Phi \in \mathbb{R}^{m\times m}$ as $\theta_{r,s}=\rho \delta_{r,s} +(1-\rho)/2^m$, where $\delta_{r,s}$ is Kronecker delta \cite{wang1995markov}. $m$-ary searching is equivalent to dividing the $L$-bit sequence to $2^{m-1}$ blocks and the number of possible initialization is $\frac{n}{2^m}$. Thus, $\P(\text{MLTS succeeds} ~|~ \text{$K_D$ established})$ is given as
\begin{eqnarray}
&& \P(\text{MLTS succeeds} ~|~ \text{$K_D$ established})\nonumber\\
&& =  { \frac{L}{2^{m-1}} \choose 0} \theta_{1,1}^{\frac{L}{2^{m-1}}} + \cdots  + { \frac{L}{2^{m-1}} \choose 1} \theta_{1,1}^{\frac{L}{2^{m-1}} - 1} \theta_{1,2} + \cdots \nonumber\\
&&   + { \frac{L}{2^{m-1}} \choose 1} \theta_{1,1}^{\frac{L}{2^{m-1}}-1} \theta_{1m} + \cdots + { \frac{L}{2^{m-1}} \choose \frac{n}{2^m}} \theta_{1,1}^{{\frac{L}{2^{m-1}}} -\frac{n}{2^m}}\theta_{1,2}^{\frac{n}{2^m}}\nonumber \\
&&+ \cdots + { \frac{L}{2^{m-1}} \choose \frac{n}{2^m}} \theta_{1,1}^{{\frac{L}{2^{m-1}}} -\frac{n}{2^m}}\theta_{1,2}^{\frac{n}{2^m}} + \cdots \nonumber\\
&& = \sum_{s=1}^{m}\sum_{i=0}^{\frac{n}{2^m}} {\frac{L}{2^{m-1}} \choose i} \theta_{s,s}^{\frac{L}{2^{m-1}}-i} \theta_{s,\neg s}^{i} \nonumber\\
&& = \sum_{s=1}^{m}I_{\rho + \frac{1-\rho}{2^m}}\left(\frac{L}{2^{m-1}}-\frac{n}{2^m}, \frac{n}{2^m}+1\right),\nonumber
\end{eqnarray}
where $\theta_{s,\neg s}$ is the transition probability from state $s$ to the rest of states except $s$.

\section{Theoretical Results of NIST Randomness Tests}\label{Sec:TestResult}
Due to the page limitation, we give the results of other 7 different tests and ignore the proof. From NIST test suite \cite{nisttest}, we can conclude that the P-value can be calculated by Gaussian distribution: frequency test (Frequency), run test (Run) and DFT test (DFT) and Chi-square distribution: frequency test within a block test (BlockFreq), longest run of ones in a block test (LongRun), non-overlapping template matching test (Nonoverlap), approximate entropy test (AppEntropy), first order serial (1storder) and second order serial (2ndorder).

\noindent\textbf{Run:} $\P(T_{\text{run}}~\text{accepts}~H_0 ) = h_{\text{run}}(\rho, \alpha) = \frac{1}{2}\text{erf}\left(\frac{\alpha_f + \mu_f}{\sigma_f \sqrt{2}}\right)$\\$+\frac{1}{2}\text{erf}\left(\frac{\alpha_f - \mu_f }{\sigma_f\sqrt{2}}\right)$, where $\alpha_f = 2\sqrt{2L}\pi(1-\pi)$$\text{erfc}^{-1}(\alpha)$, $\mu_f = L\lambda-2L(1 - \pi)$ and $\sigma_f = \sqrt{L\lambda(1 - \lambda)}$. The definition of $\pi$ is the probability of $I_l$, and $\lambda$ is the meaning of $I_l = 1$ if the $l$th element $\neq$ the ($l$-$1$)th element; $I_l = 0$ otherwise.

\noindent\textbf{DFT:} $\P(T_{\text{DFT}}~\text{accepts}~H_0 ) = h_{\text{DFT}}(\rho, \alpha) = \frac{1}{2}\text{erf}\left(\frac{\alpha_f + \mu_f}{\sigma_f \sqrt{2}}\right)+\frac{1}{2}\text{erf}\left(\frac{\alpha_f - \mu_f }{\sigma_f\sqrt{2}}\right)$, where $\alpha_f = \text{erfc}^{-1}(\alpha)\sqrt{\P_{\text{DFT}}(1-\P_{\text{DFT}})\frac{n}{4}}$, $\mu_f = 0.95\frac{n}{2} - \P_{\text{DFT}}\frac{n}{2}$, and $\sigma_f = \sqrt{(0.95)(0.05)\frac{n}{4}}$. Accordingly, $\P_{\text{DFT}} = \P(n|S_{j}(R)|<-n\ln(0.05))$, where $S_{j}(R)$ is defined in NIST.

\noindent\textbf{BlockFreq:} The statistic is $\chi^2 (\text{obs}) = 4M\sum_{i=1}^{N}(\pi_i - \frac{1}{2})^2$. Similar to the result of Frequency, $\pi_i \sim \mathcal{N}(\frac{1}{2}, \frac{4(1+\rho)}{1-\rho})$. The probability is $\P(T_{\text{longestrun}}~\text{accepts}~H_0 ) = h_{\text{longestrun}}(\rho, \alpha ) = \text{igam}\left(\frac{N}{2}, \frac{2\text{igamc}^{-1}(N/2,\alpha)\cdot\chi^2 (\text{newobs})}{\chi^2 (\text{obs})}\right) - \text{igam}\left(\frac{N}{2},\chi^2 (\text{newobs})\right)$, where igam is incomplete gamma integral function and $\text{igamc}^{-1}$ is inverse complemented incomplete gamma integral function. obs is the statistic by calculating the number and newobs is calculated by the exact distribution $\P(\pi_i )$.

\noindent\textbf{LongRun:} The statistic is $\chi^{2}(\text{obs}) = \sum_{i=0}^{K}\frac{(v_i - N\pi_i )^2}{N\pi_i}$, where $\pi_i$ is the statistical probability of obs for $i$, $K$ and $N$ are defined in NIST test suite. The exact distribution of $v_i$ is $\P(v_i) = \xi_0 \mathbf{M}^i \mathbf{1}^T$, where $\xi_0$ is the initial $[1/2 , 1/2 , 0, \dots, 0]_{1 \times (i+1)}$, $\mathbf{1}^T$ is the transpose of the row vector $\mathbf{1} = [1,1,\dots,1]_{1 \times (i+1)}$, and the $(i+1) \times (1+1)$ matrix $\mathbf{M}$ is
\begin{equation}
\mathbf{M} = \left[
\begin{matrix}
 0          ~~~& \frac{1}{2}  ~~& \frac{1}{2}      ~~& 0             ~~\cdots    ~~& 0                   \\
 0          ~~~& \P_{00}      ~~& \P_{01}          ~~& 0             ~~\cdots    ~~& 0               \\
 0          ~~~& \P_{10}      ~~& 0                ~~& \P_{11}       ~~\cdots    ~~& 0                   \\
 \vdots     ~~~& \vdots       ~~& \vdots           ~~& \vdots        ~~\ddots    ~~& 0               \\
0           ~~~& \P_{10}      ~~& 0                ~~& 0             ~~\cdots    ~~& \P_{11}             \\
0           ~~~& \P_{10}      ~~& 0                ~~& 0             ~~\cdots    ~~& 0               \\
\end{matrix}
\right]\nonumber
\end{equation}
where $\P_{00} = \P_{11} = 1/2+\rho/2$ and $\P_{01} = \P_{10} = 1/2 - \rho/2$. The probability of longestrun is $\P(T_{\text{longestrun}}~\text{accepts}~H_0 )  = h_{\text{longestrun}}(\rho, \alpha ) = \text{igam}\left(\frac{K}{2}, \frac{2\text{igamc}^{-1}(K/2,\alpha)\cdot\chi^2 (\text{newobs})}{\chi^2 (\text{obs})}\right)
$\\$ - \text{igam}\left(\frac{K}{2},\chi^2 (\text{newobs})\right)$.

\noindent\textbf{Nonoverlap:} The statistic is $\chi^{2}(\text{obs}) = \sum_{i=1}^{N}\frac{(W_i - \mu)^2}{\sigma^2}$, where $\mu = (M-m+1)/2^m$ and $\sigma^2 = M\left(\frac{1}{2^m} - \frac{2m-1}{2^{2m}}\right)$. Now, we will compute the exact distribution of $W_i$. Let $W_i$ be a template with binary numbers from $\mathcal{A} = \{0,1\}$, and define the indicator variable $I_a (W_i )$ of the appearance of $W_i$ at the position $a$ is $I_a (W_i ) = I[X_{a-m+1}=a_1 , \dots, X_a = a_m ]$ with the expectation $\eta = \pi(a_1 )\prod_{t=1}^{m-1}\P(a_t , a_{t+1})$. $\beta = (M - m + 1)y$, where $y$ is the solution of equation $ye = \eta$. Because the Markovian hypothesis and the combinatorial structure of the problem requires some more notations in addition, $e = 1 + \sum_{t=1}^{m-1}\epsilon(t)C(t)$, where $\epsilon(t) = 1$, if there is an overlap of length $t$ two $W$s; $\epsilon(t) = 0$, otherwise. $C(t) = \P(a_{m}, a_{t + 1})$, if $t = m - 1$; $\P(a_{m}, a_{t + 1})\prod_{l = t + 1}^{m - 1}\P(a_{l}, a _{l + 1})$, if $t < m - 1$.

The quantity $C(t)$ can be thought of as being the probability of observing the $m-t$ last letters of $W$ successively. Now, $W_i$ follows by Poisson distribution $W_i \sim Po(\eta)$. Thus, $\P(T_{\text{nonoverlapping}}~\text{accepts}~H_0 ) = h_{\text{nonoverlapping}}(\rho, \alpha ) = $\\$\text{igam}\left(\frac{N}{2}, \frac{2\text{igamc}^{-1}(K/2,\alpha)\cdot\chi^2 (\text{newobs})}{\chi^2 (\text{obs})}\right) - \text{igam}\left(\frac{N}{2},\chi^2 (\text{newobs})\right)$.

\noindent\textbf{AppEntropy:} The statistic is $\chi^{2}(\text{obs}) = 2N[\log 2 - ApEn(m)]$, where $ApEn = \phi^{(m)} - \phi^{(m+1)}$, $m$ is the overlapping block size and $\phi^{(m)} = \sum_{i=1}^{2^m}\frac{\text{freq of block}~i}{N}\times \log (\frac{\text{freq of block}~i}{N})$. Based on Markov dependent random variables, we define $\P_{i}^{(m)} = \P(U_{i_1})\P(U_{i_2}|U_{i_1})\cdot\P(U_{i_m}|U_{i_{m-1}})$, where the conditional probability can be calculated by transition matrix with $\rho$ in \cite{lindqvist1978note}. Now, we can define the new AppEntropy as $\chi^2 (\text{newobs})=2N[\log 2 - \tilde{\phi^{(m)}} + \tilde{\phi^{(m+1)}}]$, where $\tilde{\phi^{(m)}} = \sum_{i=1}^{2^m}\P_i^{(m)}\log \P_i^{(m)}$. Thus, we obtain that $\P(T_{\text{AppEntropy}}~\text{accepts}~H_0 )=h_{\text{AppEntropy}}(\rho, \alpha)=\text{igam}\left(2^{m-1}, \frac{\text{igamc}^{-1}\left(2^{m-1},\alpha\right)\cdot\chi^2 (\text{newobs})}{chi^2 (\text{obs})}\right) - \text{igam}\left(2^{m-1}, \frac{\chi^2 (\text{newobs})}{2}\right)$.

\noindent\textbf{1storder:} For serial test, $v_{i_m}$, $v_{i_{m-1}}$ and $v_{i_{m-2}}$ denotes the $m$-bit, $m-1$-bit and $m-2$-bit matching pattern. $U$ is each random variable from source emitting in one matching block. For Markov dependent random variables, we can obtain each matching probability of different matching pattern $\P_{i_m}(v_{i_m} = U_j ) = \P(U_1 )\times \P(U_2 | U_1 ) \times \cdots \times \P(U_m | U_{m-1})$, and the expectation is $\E_{i_m}(v_{i_m} = U_j ) = n\times\P(U_1 )\times \cdots\times\P(U_m | U_{m-1})$. The statistic of $\psi^2_m = \sum_{i_m = 1}^{2^m}\frac{(v_{i_m} - \E_{i_m})^2}{\E_{i_m}}$. Then, we compute $\Delta\hat{\psi}_{m}^{2} = \psi_{m}^{2} - \psi_{m - 1}^{2}$ and $\Delta^{2}\hat{\psi}_{m}^{2} = \psi_{m}^{2} - 2\psi_{m - 1}^{2} + \psi_{m - 2}^{2}$. The probability of first order serial is $\P(T_{\text{firstorder}}~\text{accepts}~H_0) = h_{\text{firstorder}}(\rho, \alpha)
= \text{igam}\left(2^{m - 2}, \frac{2\text{igamc}^{-1}(2^{m - 2}, \alpha)\cdot\Delta\hat{\psi}_{m}^{2}}{\Delta\psi_{m}^{2}}\right)
 - \text{igam}\left(2^{m - 2}, \Delta\hat{\psi}_{m}^{2}\right)$.

\noindent\textbf{2ndorder:} Similarly, the probability of 2ndorder is \\
$\P(T_{\text{secondorder}}~\text{accepts}~H_0) = h_{\text{secondorder}}(\rho, \alpha) = $\\$\text{igam}\left(2^{m - 3}, \frac{2\text{igamc}^{-1}(2^{m - 3}, \alpha)\cdot \Delta^{2}\hat{\psi}_{m}^{2}}{\Delta^{2}\psi_{m}^{2}}\right) - \text{igam}\left(2^{m - 3}, \Delta^{2}\hat{\psi}_{m}^{2}\right)$.



\bibliographystyle{IEEEtran}
\bibliography{keyattack}
\end{document}